\documentclass{elsart5p}
\journal{Physica D}
\usepackage{lineno}
\usepackage{amsfonts}
\usepackage{amssymb}
\usepackage{amsmath}
\usepackage{epsfig}
\usepackage{graphicx}
\usepackage{color}

\def\rootfig{./2D-ps/}

\begin{document}
%\linenumbers

\begin{frontmatter}
\title{Multistable Solitons in Higher-Dimensional Cubic-Quintic
Nonlinear Schr\"{o}dinger Lattices}

\author[germany]{C. Chong\corauthref{cor1}}
\corauth[cor1]{Corresponding author},
\ead[url]{http://www.iadm.uni-stuttgart.de/LstAnaMod/Chong/home.html}
\author[sdsu]{R.\ Carretero-Gonz\'alez},
\author[israel]{B.A.\ Malomed}, and
\author[pgk]{P.G.\ Kevrekidis}

\address[germany]{%
Institut f\"ur Analysis, Dynamik und Modellierung,
Universit\"at Stuttgart, Stuttgart 70178, Germany
}
\address[sdsu]{%
Nonlinear Dynamical Systems Group{$^1$},
%\thanksref{nlds}
Computational Sciences Research Center, and\\
Department of Mathematics and Statistics,
San Diego State University, San Diego,
CA 92182-7720, USA
}
\thanks[nlds]{{\tt URL:} http://nlds.sdsu.edu/}
\address[israel]{%
Department of Interdisciplinary Studies, Faculty of Engineering,
Tel Aviv University, Tel Aviv 69978, Israel
}
\address[pgk]{%
Department of Mathematics and Statistics, University of
Massachusetts, Amherst MA 01003-4515, USA}

%\author{C.\ Chong\footnote{\texttt{E-MAIL:} christopher.chong@mathematik.uni-stuttgart.de\\
%\texttt{\phantom{$^*$}URL:} http://www.iadm.uni-stuttgart.de/LstAnaMod/Chong/home.html}}
%\affiliation{Institut f\"ur Analysis, Dynamik und Modellierung, Universit\"at Stuttgart,
%Stuttgart 70178, Germany}
%\author{R.\ Carretero-Gonz\'alez}
%\affiliation{Nonlinear Dynamical Systems Group,\footnote{\texttt{URL:} http://nlds.sdsu.edu/%
%} Computational Sciences Research Center, and Department of Mathematics and
%Statistics, San Diego State University, San Diego, CA 92182-7720, USA}
%\author{B. A.\ Malomed}
%\affiliation{Department of Physical Electronics, School of Electrical Engineering,
%Faculty of Engineering, Tel Aviv University, Tel Aviv 69978, Israel}
%\author{P. G.\ Kevrekidis}
%\affiliation{Department of Mathematics and Statistics, University of Massachusetts,
%Amherst MA 01003-4515, USA}

%email address: christopher.chong@mathematik.uni-stuttgart.de
%web page: http://www.iadm.uni-stuttgart.de/LstAnaMod/Chong/

\begin{abstract}
We study the existence, stability, and mobility of fundamental
discrete solitons in two- and three-dimensional nonlinear
Schr\"{o}dinger lattices with a combination of cubic self-focusing
and quintic self-defocusing onsite nonlinearities. Several species
of stationary solutions are constructed, and bifurcations linking
their families are investigated using parameter continuation starting
from the anti-continuum limit, and also with the help of a
variational approximation. In particular, a species of hybrid
solitons, intermediate between the site- and bond-centered types
of the localized states (with no counterpart in the 1D model), 
is analyzed in 2D and 3D lattices.
%, while its
%counterpart in the 1D model does not exist. 
We also discuss the
mobility of multi-dimensional discrete solitons that
can be set in motion by lending them kinetic energy exceeding the
appropriately crafted Peierls-Nabarro barrier; however,
they eventually come to a halt,
due to radiation loss.
\end{abstract}

\maketitle

%%use optional labels to link authors explicitly to addresses:
%%\author[label1,label2]{}
%%\address[label1]{}
%%\address[label2]{}


\end{frontmatter}

%%%%%%%%%%%%%%%%%%%%%%%%%%%%%%%%%%%%%%%%%%%%%%%%%%%%%%%%%%%%%%%%%%%%%%%%%%%%%%%
\section{Introduction}

A large number of models relevant to various fields of physics is based on
discrete nonlinear Schr\"{o}dinger (DNLS) equations \cite{Panos}. A
realization of the one-dimensional (1D) DNLS model in arrays of parallel
optical waveguides was predicted in Ref.~\cite{Demetri}, and later
demonstrated experimentally, using an array mounted on a common substrate
\cite{Silberberg}. Multi-channel waveguiding systems can also be created as
photonic lattices in bulk photorefractive crystals \cite{Moti}. Discrete
solitons are fundamental self-supporting modes in the DNLS system \cite%
{Panos}. The mobility \cite{Boulder,Papa} and collisions \cite%
{Papa,interaction} of discrete solitons have been studied in 1D
systems of the DNLS type with the simplest self-focusing cubic
(Kerr) nonlinearity. The DNLS equation with the cubic onsite
nonlinearity is also a relevant model for the Bose-Einstein
condensate (BEC) trapped in deep optical lattices \cite{discrete}.

A more general discrete cubic nonlinearity appears in the Salerno model \cite%
{Salerno}, which combines the onsite cubic terms and nonlinear coupling
between adjacent sites. A modification of the Salerno model, with opposite
signs in front of the onsite and inter-site cubic terms, makes it possible
to study the competition between self-focusing and defocusing discrete
nonlinearities. This has been done in both 1D \cite{Zaragoza1} and 2D \cite%
{Zaragoza2} settings.

Lattice models with saturable onsite nonlinear terms have been studied too. The
first model of that type was introduced by Vinetskii and Kukhtarev in 1975
\cite{Russian}. Bright solitons in this model were predicted in 1D \cite%
{LjupcoSandra} and 2D \cite{Sweden} geometries. Lattice solitons
supported by saturable self-defocusing nonlinearity were created in
an experiment conducted in an array of optical waveguides built in a
photovoltaic medium \cite{photovoltaic}. Dark discrete solitons were
also considered experimentally \cite{photovoltaic2} and theoretically
\cite{fitrakis} in the latter model.

The experimental observation of optical nonlinearities that may be fitted by
a combination of self-focusing cubic and self-defocusing quintic terms \cite%
{CQoptical} suggests to study the dynamics of solitons in the NLS
equation with cubic-quintic (CQ) nonlinearity. A family of stable
exact soliton
solutions to the 1D continuum NLS equation of this type is well known \cite%
{Bulgaria}. The possibility to build an array of parallel waveguides
using optical materials with the CQ nonlinearity lends relevance to
the consideration of the DNLS equation with the onsite nonlinearity
of the CQ type. In particular, this DNLS equation arises as a limit
case of the continuum CQ-NLS equation which includes a periodic
potential in the form of periodic array of rectangular channels,
i.e., the Kronig-Penney lattice. Families of stable %families of
bright solitons were found in 1D \cite{Radik1} and 2D \cite{Radik2}
versions of the latter model (the latter one with a ``checkerboard"
2D potential supports both fundamental and vortical solitons).

The findings of a CQ-DNLS model may also be relevant to the case  of a
self-attractive BECs confined in a 2D plane by a ``pancake"-shaped
trap combined with a sufficiently strong 2D optical-lattice potential
(although quantum effects such as a superfluid to Mott insulator
transition are also relevant in the latter \cite{bloch}). The
condensate trapped in each individual potential well of this
configuration is described by the Gross-Pitaevskii equation with an
extra self-attractive quintic term which accounts for the deviation
of the well's shape from one-dimensionality \cite{Shlyap}. The
tunneling of atoms between adjacent potential wells in this setting
is approximated by the linear coupling between sites of the respective
lattice.

The simplest stationary bright solitons, of the \textit{unstaggered}
type (without spatial oscillations in the solitons' tails), have
been studied in the 1D version of the CQ-DNLS model in 
Ref.~\cite{Ricardo}. It was demonstrated that this class of solitons
includes infinitely many families with distinct symmetries. The
stability of the basic families was analyzed, and bifurcations
between them were explored in a numerical form, and by means of a
variational approximation (VA). Dark solitons in the same model were
recently studied \cite{Belgrade1} and, in another very recent
work, staggered 1D bright solitons as well as the mobility of
unstaggered ones have been investigated \cite{Belgrade2}.

The aim of the present work is to study the existence, stability, and
mobility of bright discrete solitons in two- and three-dimensional (2D and
3D) NLS lattices with the nonlinearity of the CQ type. As suggested by the
previous works, especially Ref.~\cite{Ricardo}, the competition of the
self-focusing cubic and self-defocusing quintic nonlinearities in the
setting of the discrete model may readily give rise to multi-stability of
discrete solitons, which is not possible in the ordinary cubic DNLS model
\cite{cubic}, nor in the discrete CQ model where both nonlinear terms are
self-focusing \cite{chongthesis}. In addition to that, one may expect that
the CQ model shares many features with those including saturable
nonlinearity \cite{Melvin,Sweden}, such as enhanced mobility of
multidimensional discrete solitons (as mentioned above, mobile discrete
solitons can be readily found in the 1D CQ-DNLS equation \cite{Belgrade2}).
%However, the saturable model does not generate the soliton multistability,
%as it does not include competing nonlinearities.

The paper is organized as follows: in the next section, we introduce the
model and outline the method used to construct the multi-dimensional
discrete solitons. In Sec.~\ref{Sec:bif}, we focus on stability and
existence regions for 2D discrete solitons, and the respective bifurcations.
Mobility of the 2D solitons on the lattice is studied in Sec.~\ref%
{Sec:mobile}. Section~\ref{Sec:3D} reports extensions of these results to 3D
latices. In Sec.~\ref{Sec:VA} we report analytical results obtained by means
of a VA, and Sec.~\ref{Sec:conclusions} concludes the paper.

\section{The model \label{Sec:model}}

In dimensionless form, the 2D DNLS equation with the onsite nonlinearity of
the CQ type has the following form:
\begin{equation}
i\dot{\psi}_{n,m}+C\Delta ^{(2)}\psi _{n,m}+2|\psi _{n,m}|^{2}\psi
_{n,m}-|\psi _{n,m}|^{4}\psi _{n,m}=0,  \label{2DCQDNLS}
\end{equation}%
where $\psi _{n,m}$ is the complex field at site \{$n,m$\} (the amplitude of
the electromagnetic field in an optical fiber, or local mean-field wave
function in BEC), $\dot{\psi}\equiv d\psi /dt$, and $C>0$ is the coupling
constant of the lattice model. We assume an isotropic medium, hence the
discrete Laplacian is taken as
\begin{equation}
\Delta ^{(2)}\psi _{n,m}\equiv \psi _{n+1,m}+\psi _{n-1,m}+\psi
_{n,m+1}+\psi _{n,m-1}-4\psi _{n,m}.
\end{equation}%
The CQ nonlinearity is represented by the last two terms in Eq.~(\ref%
{2DCQDNLS}).

Equation (\ref{2DCQDNLS}) conserves two dynamical invariants: norm (or
power, in terms of optics),
\begin{equation}
M=\sum_{n,m}\left\vert \psi _{n,m}\right\vert ^{2},  \label{M}
\end{equation}%
and energy (Hamiltonian),
%%%%%%%%%%%%%%%%%%%%%%%%%%%%%%%%%
% This Hamiltonian is derived the standard way,
% which differs in sign from the one used in the
% 1D paper.
%%%%%%%%%%%%%%%%%%%%%%%%%%%%%%%%%%
\begin{eqnarray}
H &=& \sum_{n,m}  
\left[ C (|\psi _{n+1,m}-\psi _{n,m}|^2 + |\psi _{n,m+1}-\psi _{n,m}|^2) \right.
\notag \\[1ex]
&&\phantom{\sum_{n,m}}\left. - |\psi _{n,m}|^{4}+ \frac{1}{3} |\psi _{n,m}|^{6}\right].
\label{Hamil}
\end{eqnarray}%
The conserved quantities play an important role in the analysis of the mobility
of discrete solitons, see Sec.~\ref{Sec:mobile} below.

Steady state solutions are sought for in the usual form, $\psi
_{n,m}=u_{n,m}\exp (-i\mu t)$, where $\ \mu $ is the real frequency, and
the real stationary lattice field $u_{n,m}$ satisfies the following discrete
equation:
\begin{equation}
\mu u_{n,m}+C\Delta ^{(2)}u_{n,m}+2u_{n,m}^{3}-u_{n,m}^{5}=0.
\label{CQDNLS-sta}
\end{equation}%
More general solutions carrying topological charge, for which stationary
field $u_{n,m}$ is complex, fall outside of the scope of the present work,
and will be considered elsewhere.

In one dimension, bright-soliton solutions of Eq.~(\ref{CQDNLS-sta}) can be
found as homoclinic orbits of the corresponding two-dimensional discrete map
\cite{ref:multibreathers}. This technique was used to construct 1D soliton
solutions to the CQ-DNLS model in Ref.~\cite{Ricardo}. Since this method is
not available in higher dimensions, we construct the solutions starting
from the anti-continuum limit, $C\rightarrow 0$, and perform
parameter continuation to $C>0$. A multidimensional version of the VA can
also be used to construct solutions for small values of $C$, see Sec.~\ref%
{Sec:VA} below.

\begin{figure}[t]
\centerline{
 \epsfig{file=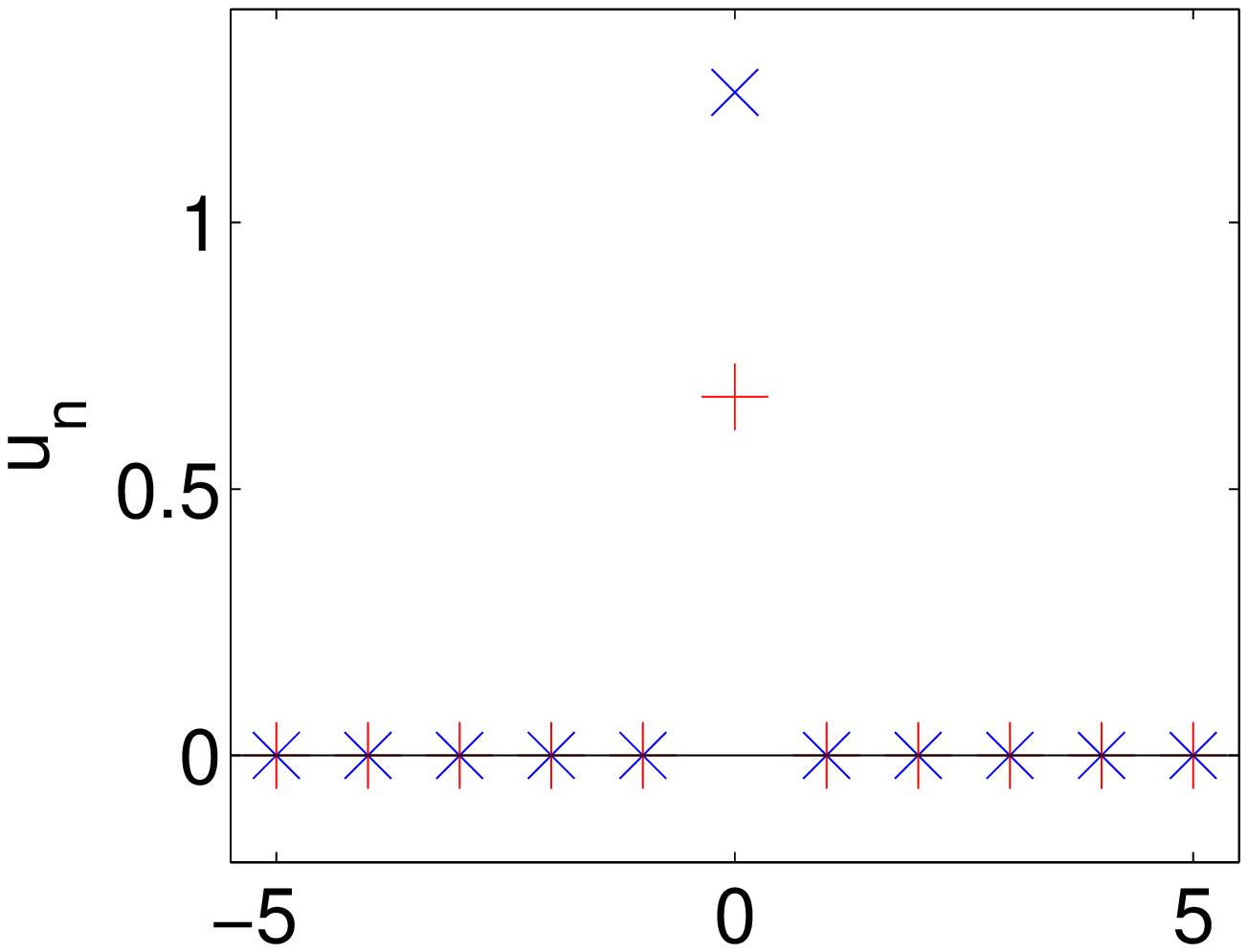,width=4.5cm,angle=0}
 \epsfig{file=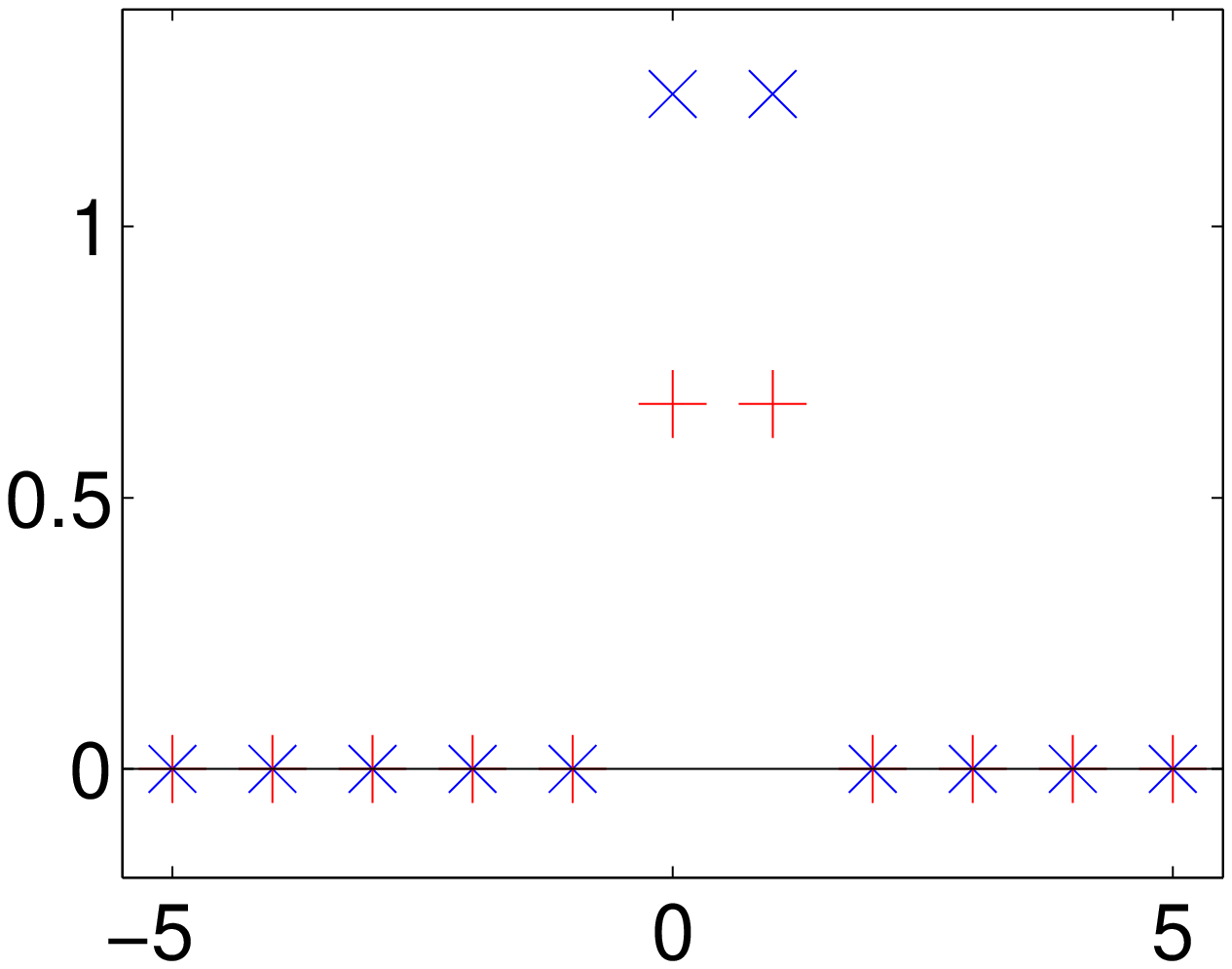,width=4.5cm,angle=0}}
\centerline{
 \epsfig{file=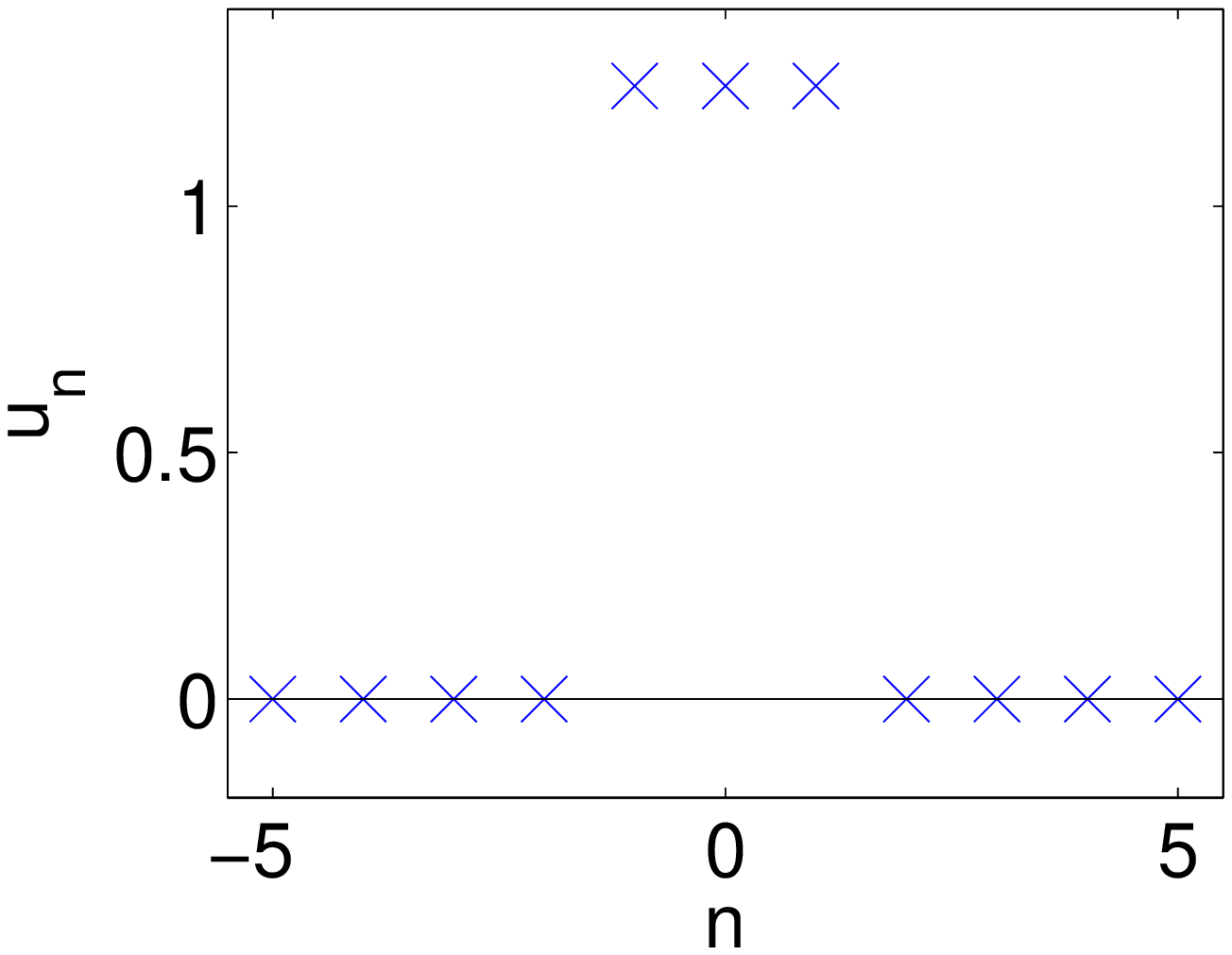,width=4.5cm,angle=0}
 \epsfig{file=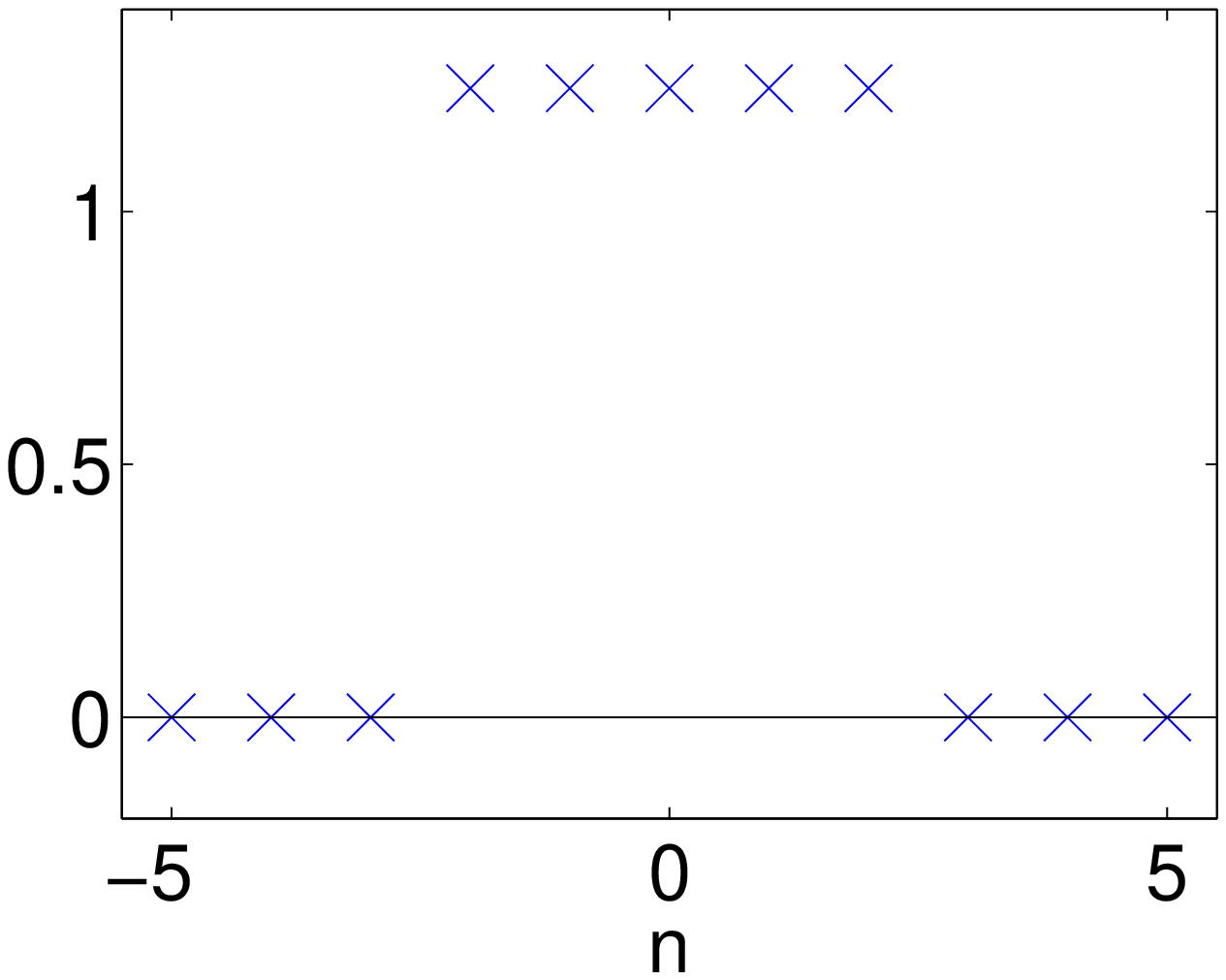,width=4.5cm,angle=0}}
\caption{(Color online) Solutions to Eq.~(\ref{2DCQDNLS}) for 
$(\mu,C)=(-0.7,0)$, which are used as seeds to find nontrivial solutions
at $C>0$ (only a 1D slice is shown, see
Fig.~\ref{2Dprofiles.ps} for profiles in two dimensions).
Top left: ``Tall" (blue cross markers) and ``short" (red plus
markers) site-centered solutions. Top right: ``Tall" and ``short"
bond-centered solutions. Bottom: wider extensions of the ``tall"
site-centered solution.} \label{seeds}
\end{figure}

\begin{figure}[t]
\centerline{
 \epsfig{file=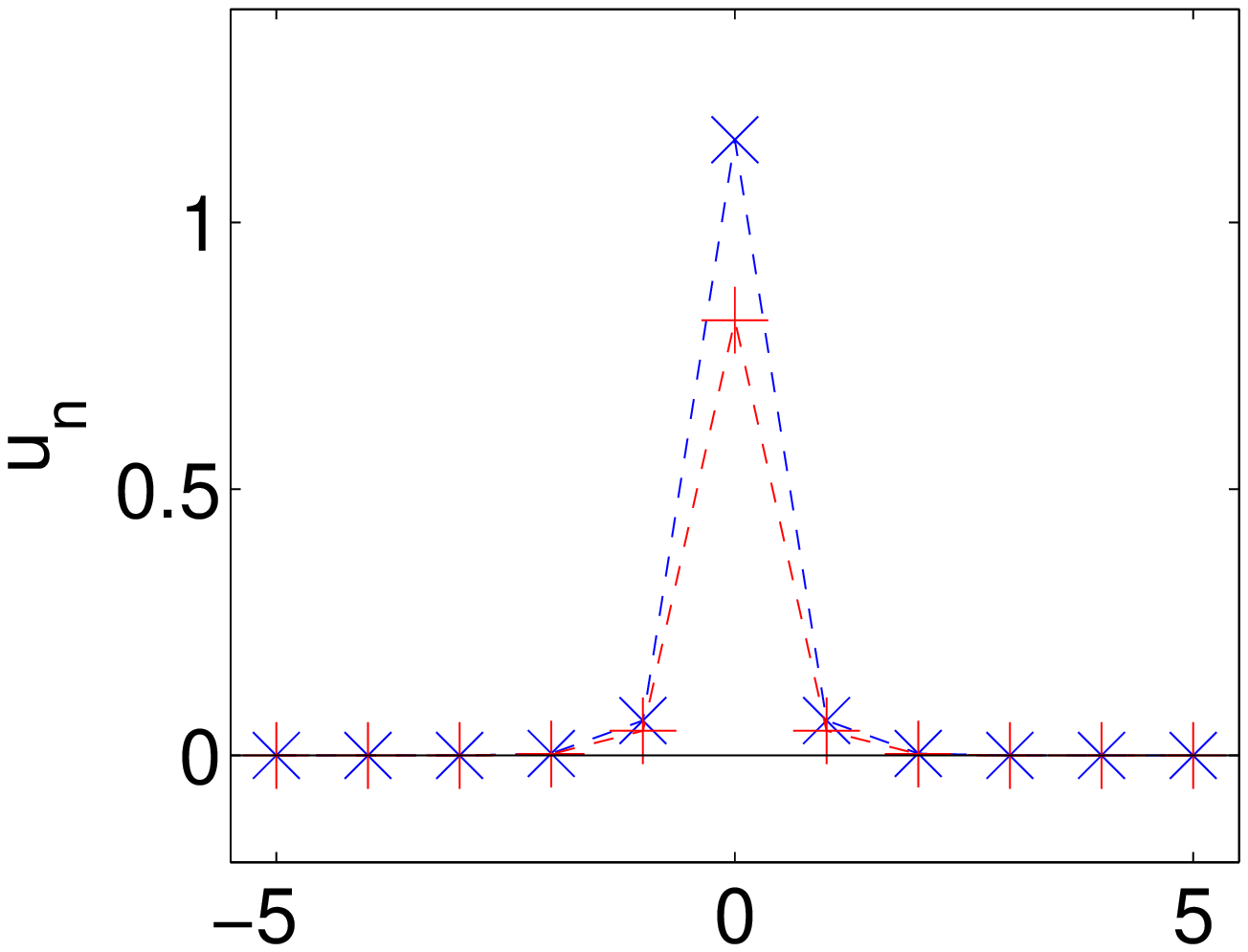,width=4.5cm,angle=0}
 \epsfig{file=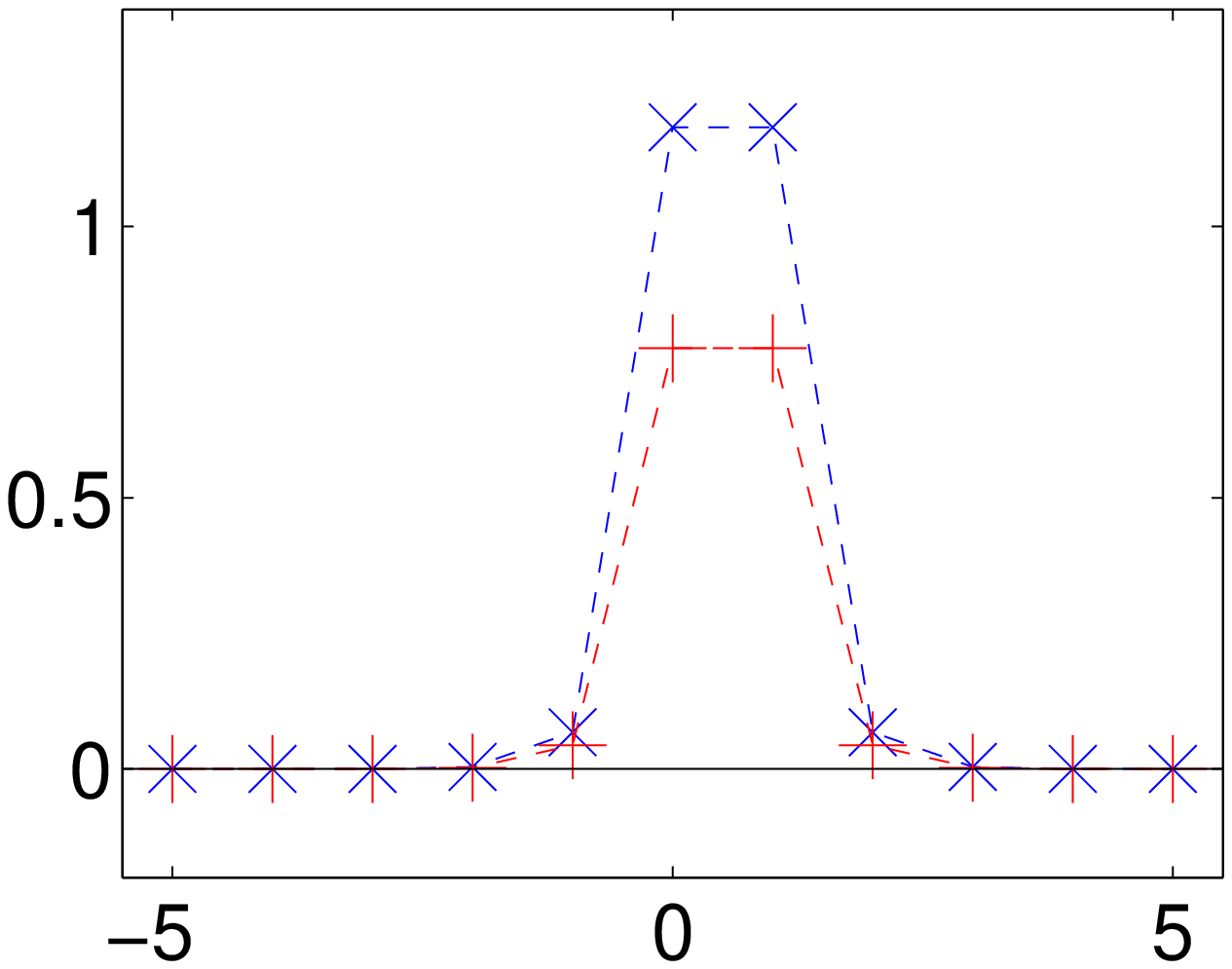,width=4.5cm,angle=0}}
\centerline{
 \epsfig{file=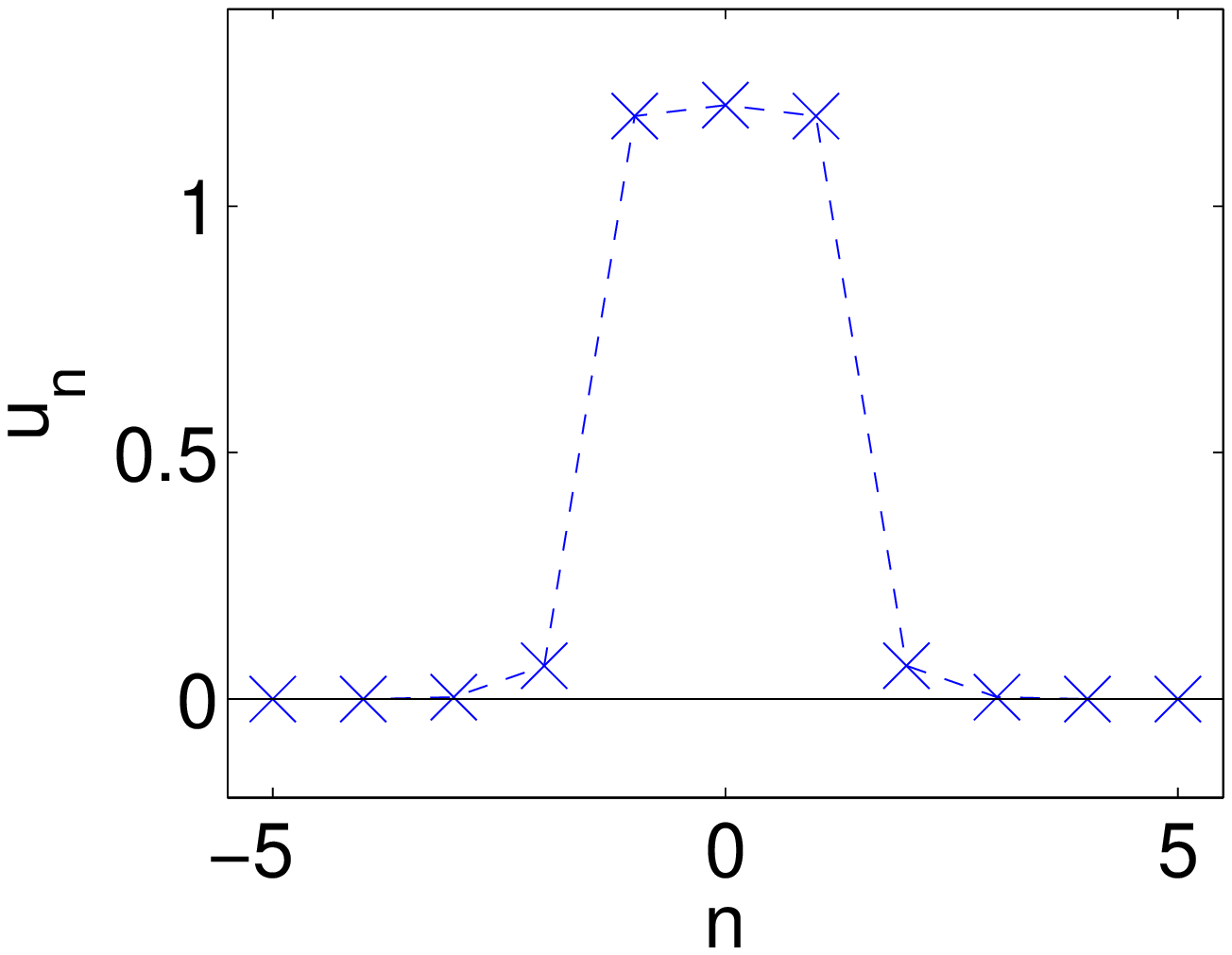,width=4.5cm,angle=0}
 \epsfig{file=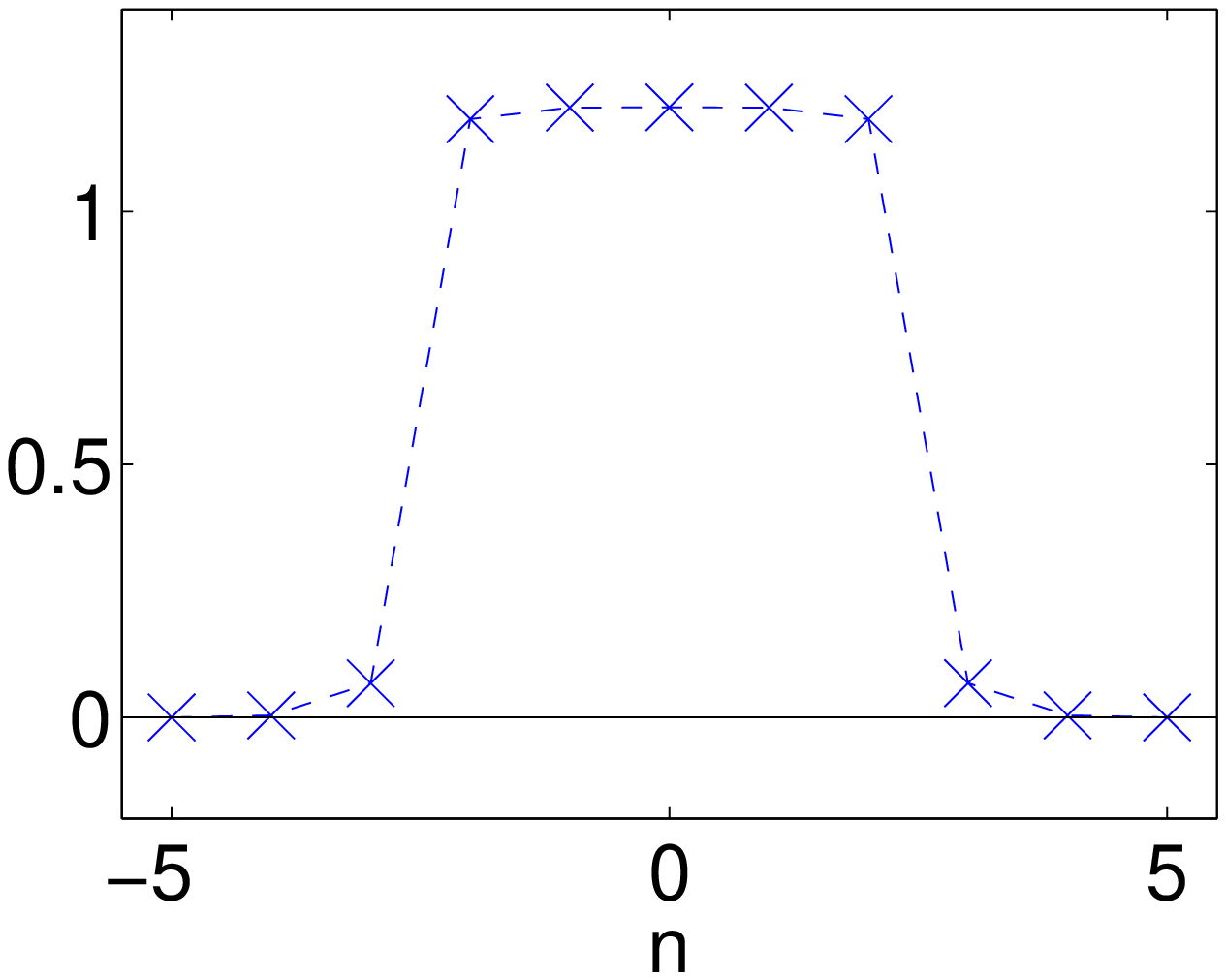,width=4.5cm,angle=0}}
\caption{(Color online) The continuation to $C=0.1$ of the solutions shown in 
Fig.~\ref{seeds}.}
\label{seeds2}
\end{figure}

In Ref.~\cite{Ricardo}, two fundamental types of solutions were
studied: site-centered and bond-centered solitons. Each family of
solutions was further subdivided into two sub-families, which
represent ``tall" and ``short" solutions for given parameter values.
Moreover, each sub-family contains, depending on the
value of $C$, wider solutions that may be built by appending extra
excited sites to the soliton. The reason for the co-existence of the
tall and short sub-families is clearly seen in the anti-continuum.
If $C=0$, Eq.~(\ref{CQDNLS-sta}) reduces to the following algebraic
equation:
\begin{equation}
\mu u_{n,m}+2u_{n,m}^{3}-u_{n,m}^{5}=0,  \label{CQDNLS-c0}
\end{equation}%
which has at most five real solutions, \textit{viz}., \ four nontrivial ones,%
\begin{equation}
u_{n,m}=\displaystyle\pm \sqrt{1\pm \sqrt{1+\mu }},  \label{fpts}
\end{equation}
and $u_{n,m}=0$ (note that these are also fixed points of the above-mentioned
discrete map in the 1D case). Obviously, Eq.~(\ref{fpts}) gives, at most,
two different non-trivial amplitudes, that may be continued to $C>0$, giving
rise to tall and short solitons respectively. To build wider solutions, one
has to consider multiple sites with the nonzero field. 
Using the $C=0$ solutions as seeds, we are able to generate a large
family of solutions in the $(\mu ,C)$ parameter plane, as shown in 
Figs.~\ref{seeds} and \ref{seeds2}. It is found that all the wide solutions tend to
disappear through saddle-node collisions between the tall and short
solutions as $C$ increases, similarly to what is the case for
the cubic DNLS problem, as discussed in Ref.~\cite{konotop}.

\begin{figure}[t]
\centerline{
 \epsfig{file=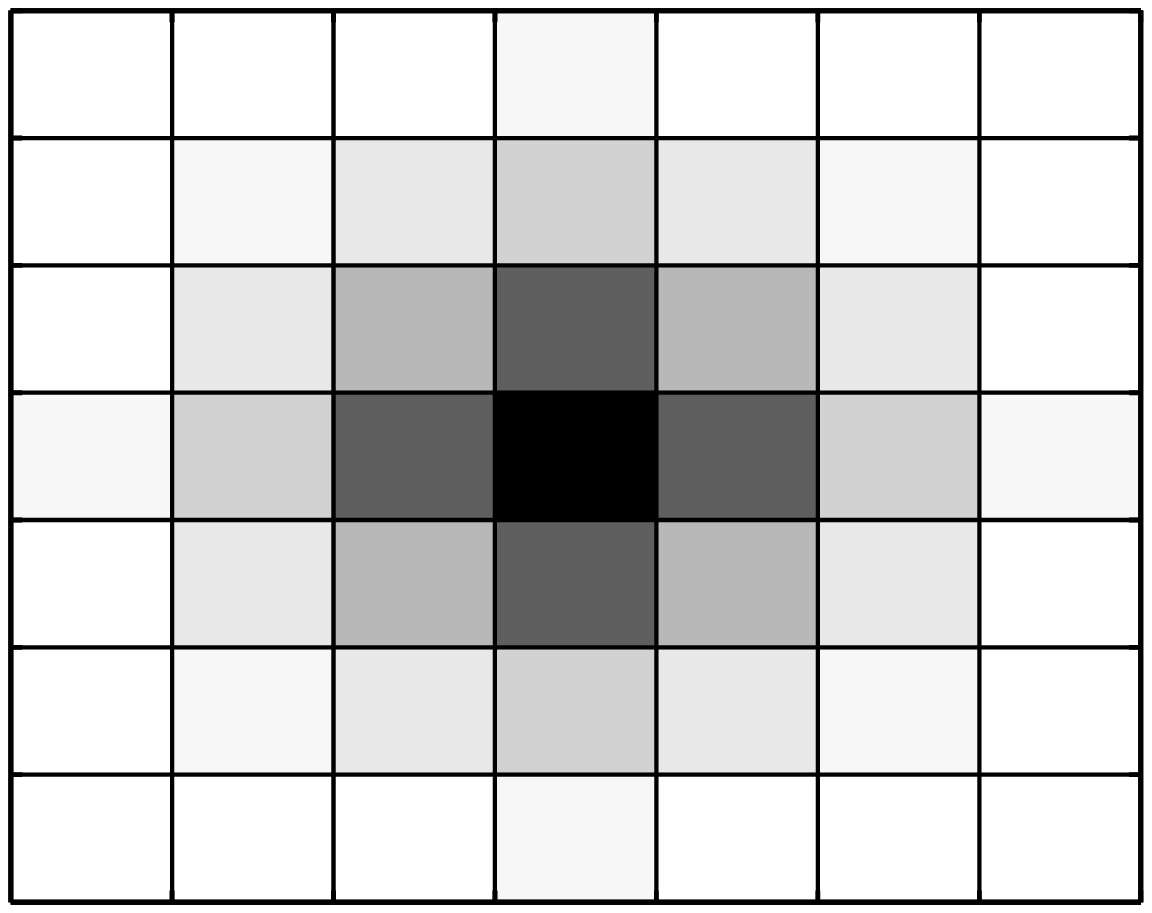, width=3.6cm,height=3.3cm,angle=0}
 \hskip-0.0cm
 \epsfig{file=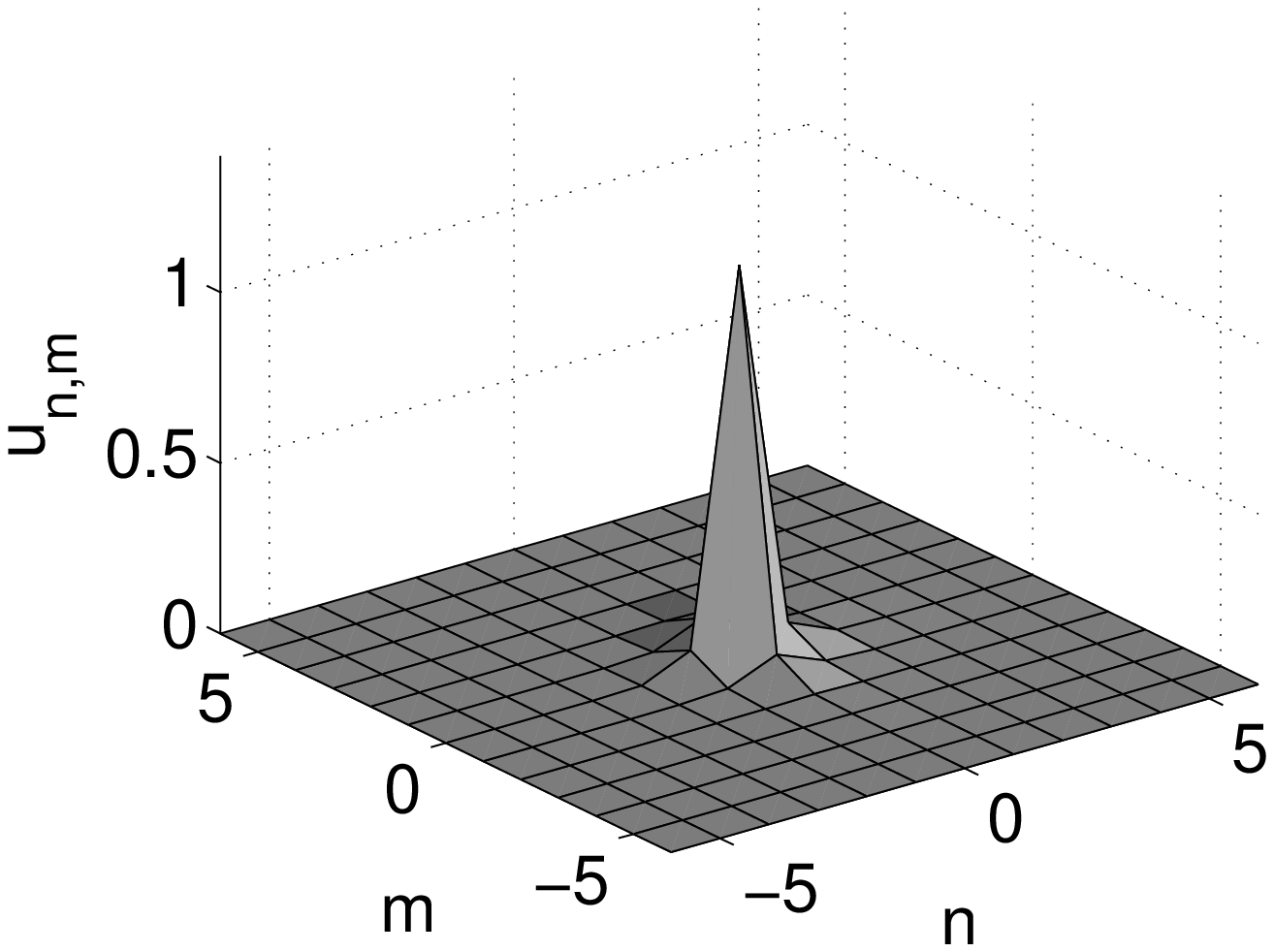,width=4.5cm,angle=0}
 }
\centerline{
 \epsfig{file=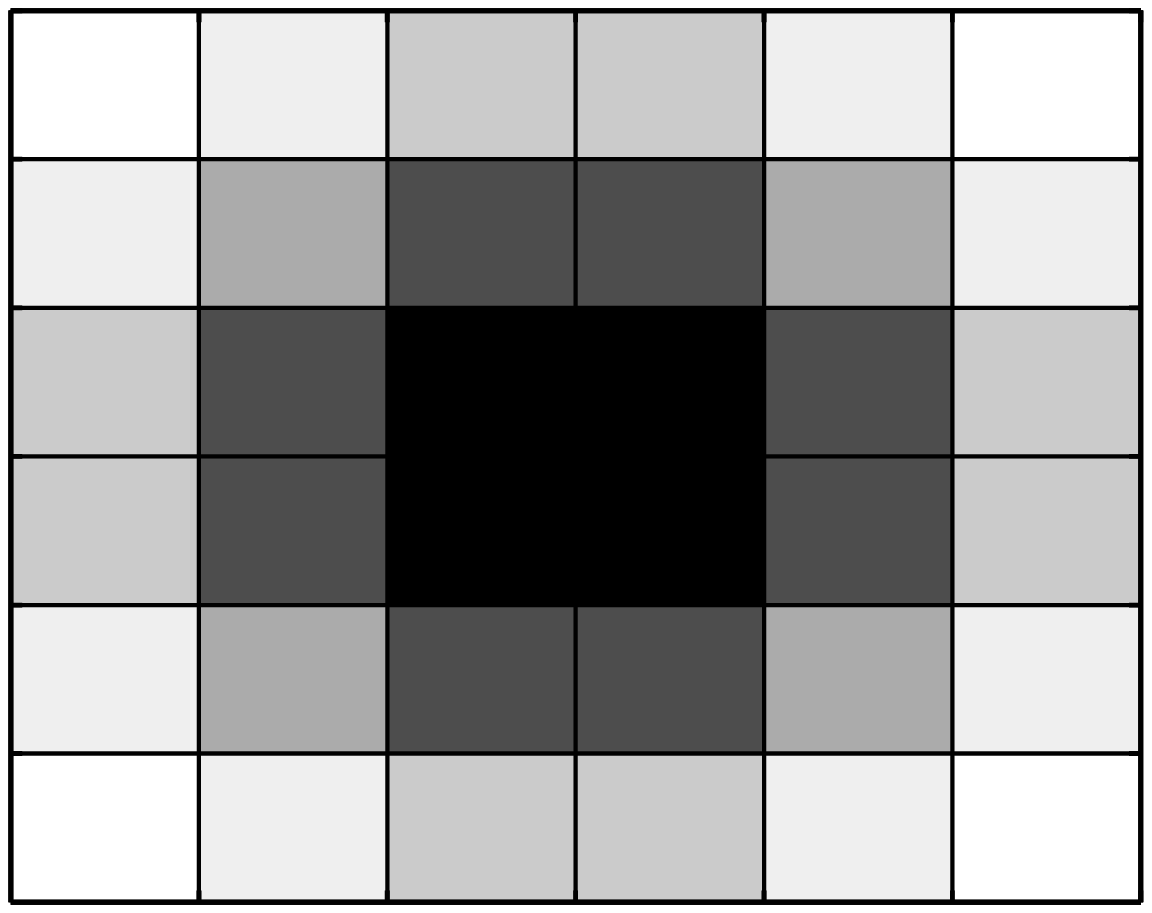, width=3.6cm,height=3.3cm,angle=0}
 \hskip-0.0cm
 \epsfig{file=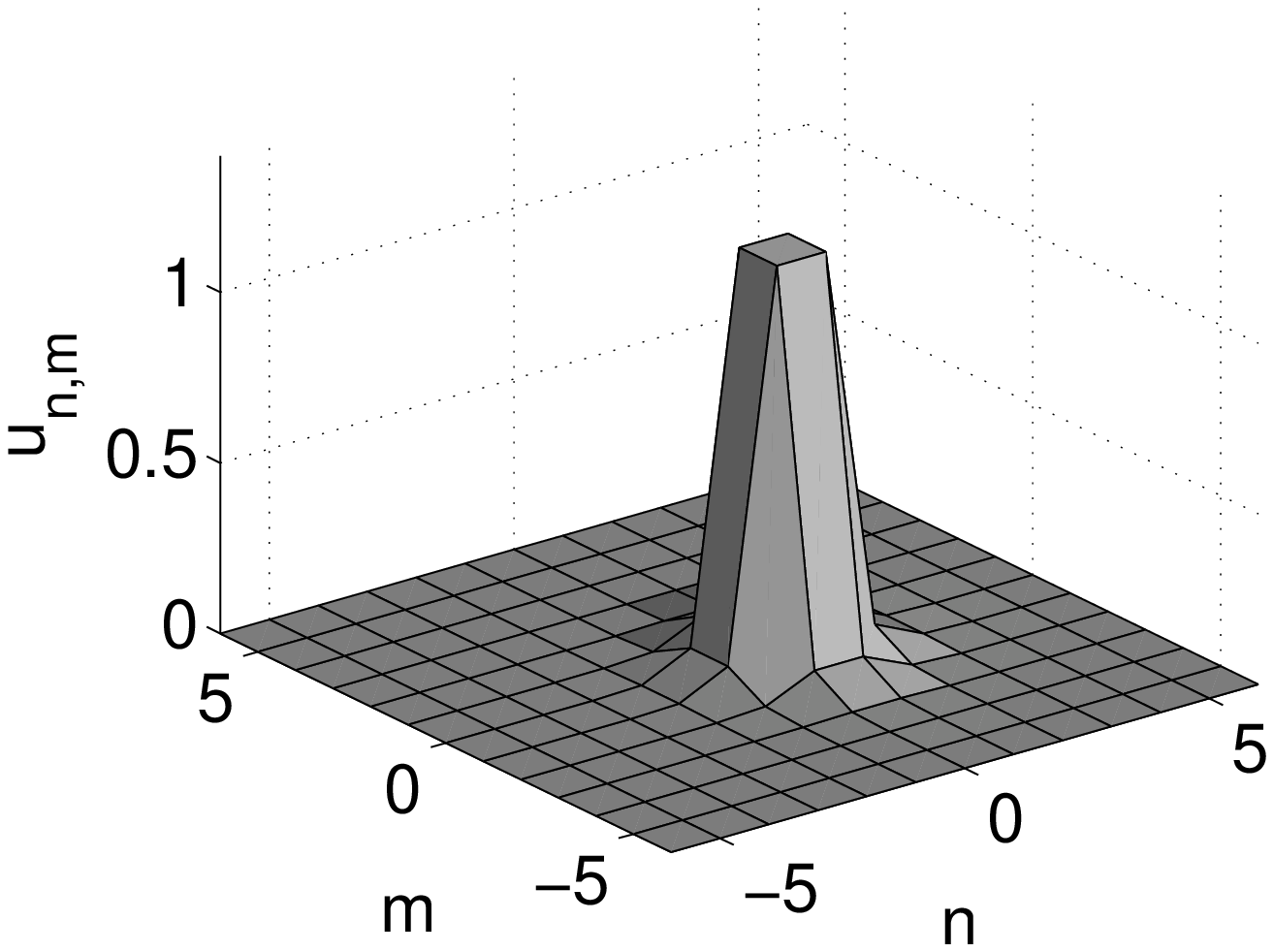,width=4.5cm,angle=0}
 }
\centerline{
 \epsfig{file=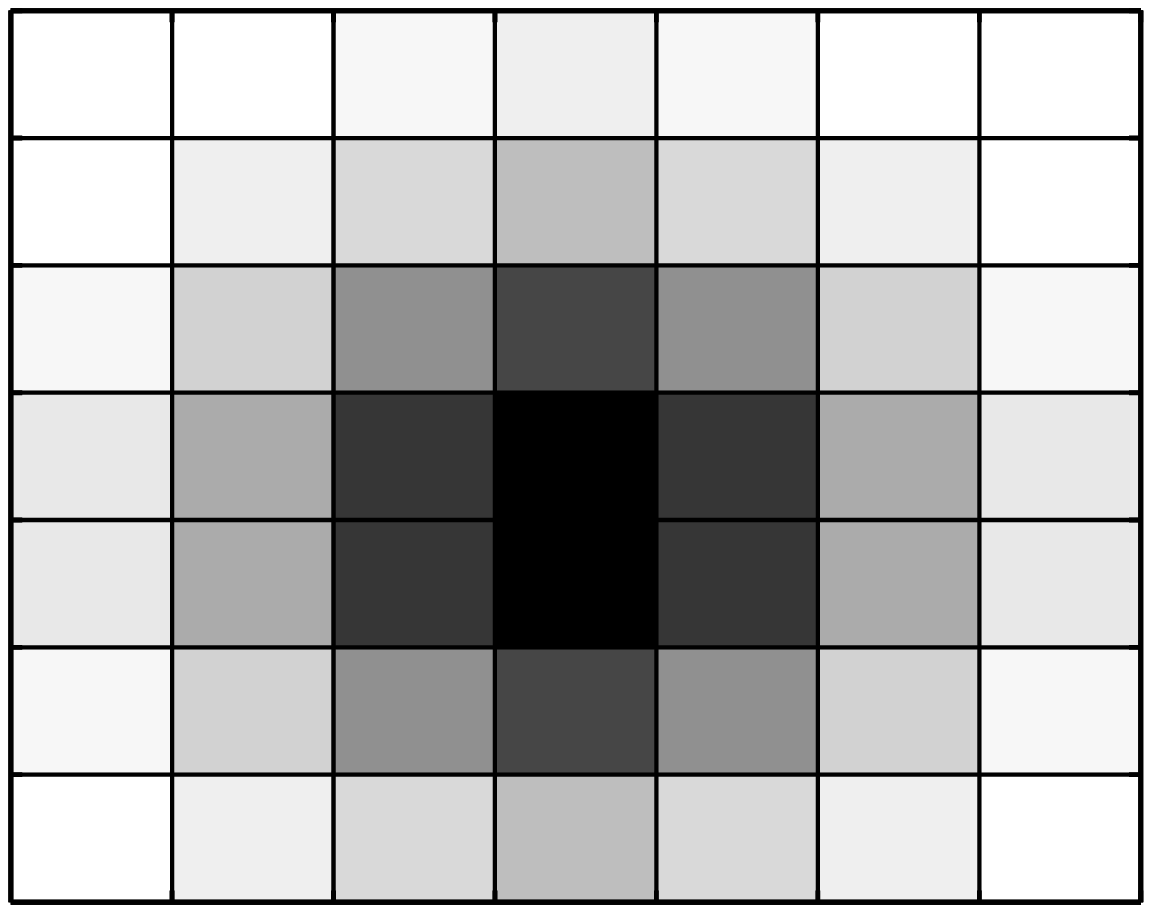,   width=3.6cm,height=3.3cm,angle=0}
 \hskip-0.0cm
 \epsfig{file=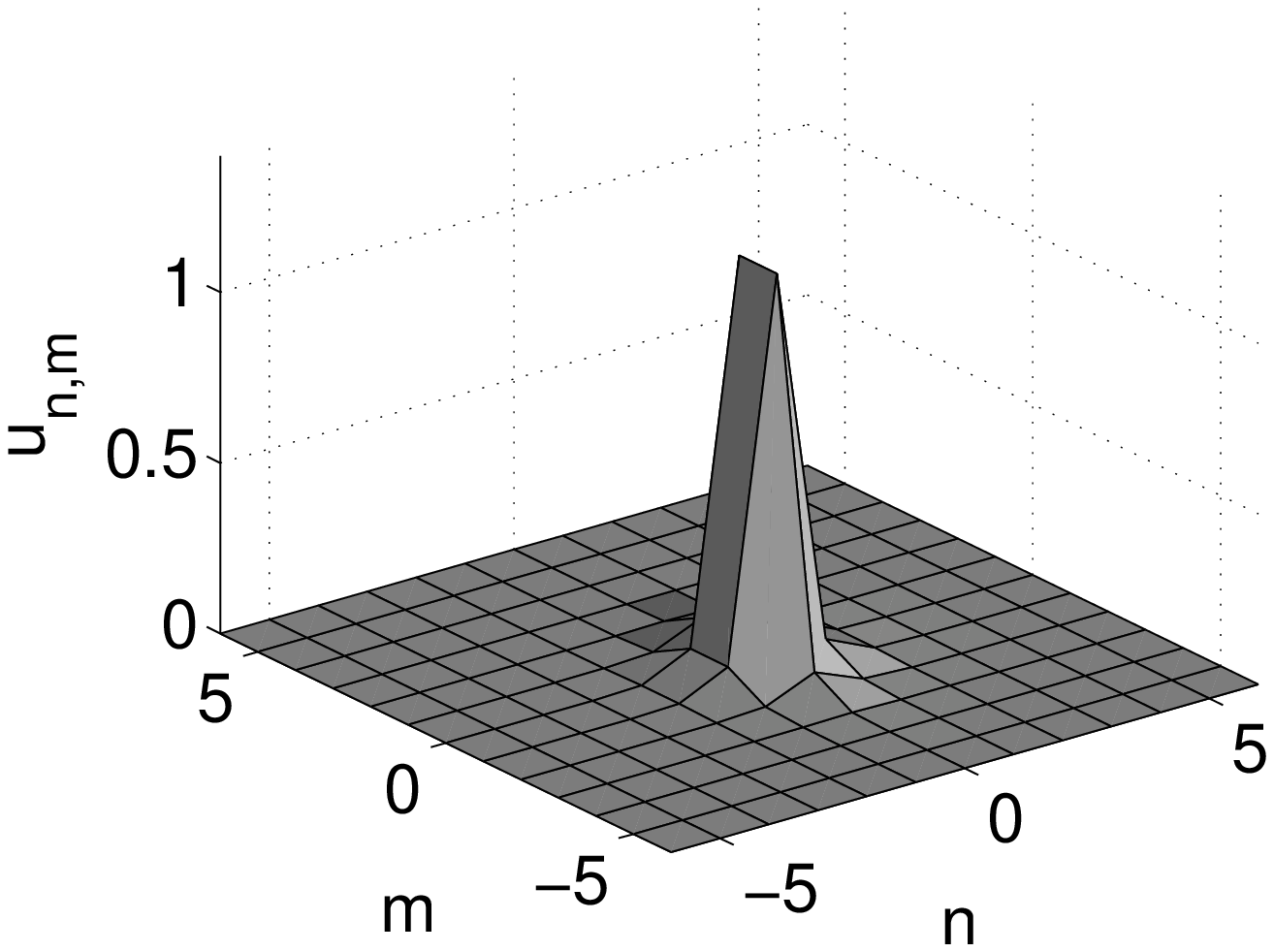,width=4.5cm,angle=0}
 }
\caption{Left (from top to bottom): Contour plots of solutions of the
site-centered, bond-centered, and hybrid types for 
$(\mu,C)=(-0.7,0.1)$. Right: The corresponding 3D plots.}
\label{2Dprofiles.ps}
\end{figure}

Another fundamental type of solution that arises in higher-dimensional
lattices is a hybrid between the site-centered and the bond-centered
solutions along the two spatial directions,
see bottom panels in Fig.~\ref{2Dprofiles.ps}. 
This type of hybrid solution was considered previously in the case
of the cubic DNLS model in Ref.~\cite{pgk2000}.
We only consider
these three symmetric types of localized states, namely the
bond-centered, site-centered, and hybrid ones (see Fig.~\ref{2Dprofiles.ps}), together with their
intermediate asymmetric counterparts (see Fig.~\ref{deform.ps}(c) for an example). 
%For the first two types, the profile along both spatial axes are identical
%(therefore they are called symmetric).
The hybrid solution admits other
natural variations, namely any combination of the various types of
bond-centered solutions along one axis and any site-centered profile along
the other. Since their behaviors are very similar, we consider only one such type
of solutions.
%Employing this simple method, one could also generate
%multi-humped solutions and dark solitons.

%%%%%%%%%%%%%%%%%%%%%%%%%%%%%%%%%%%%%%%%%%%%%%%%%%%%%%%%%%%%%%%%%%%%%%%%%%%%%%%

%combined.eps
\begin{figure*}[tbp]
\centerline{ \epsfig{file=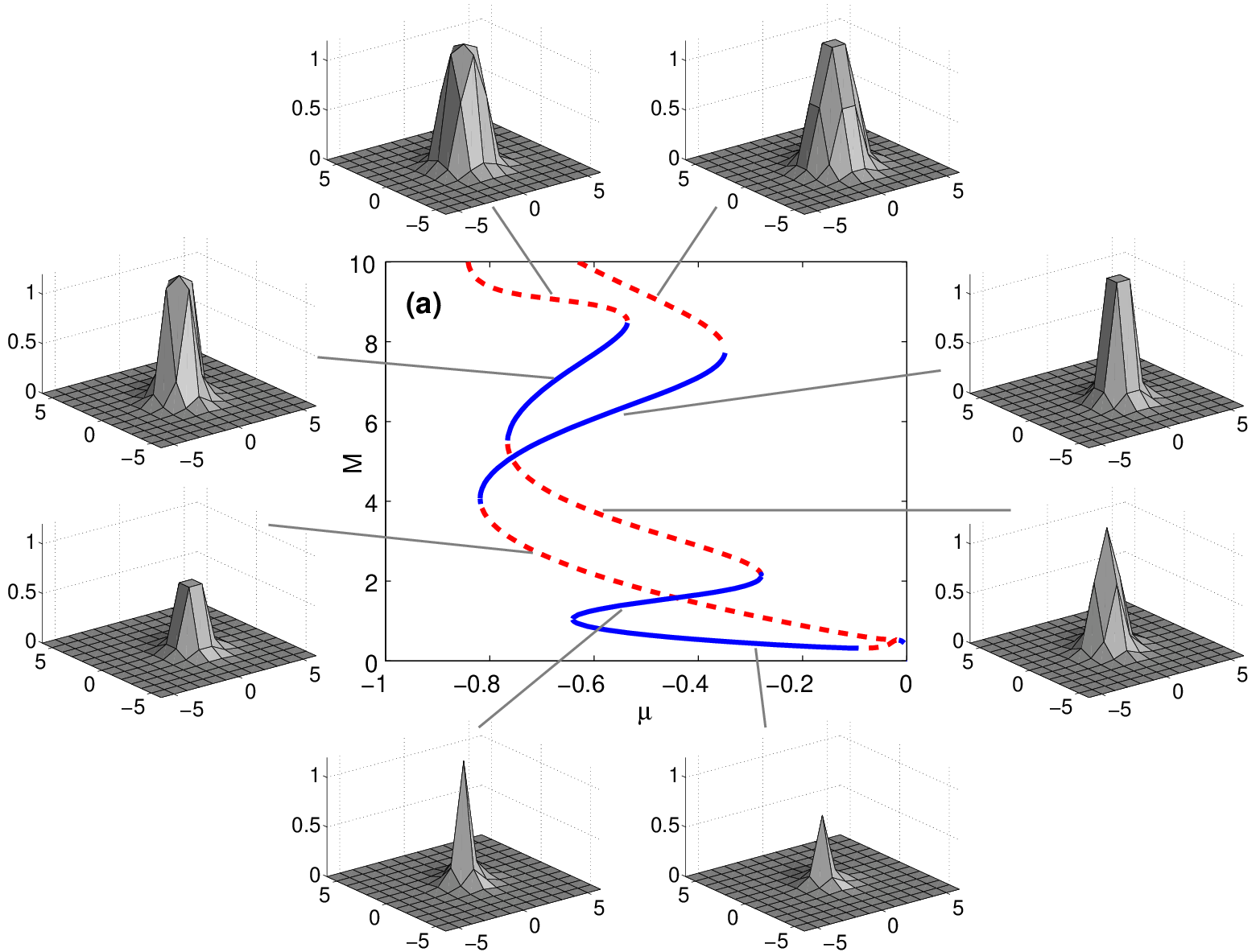,width=13cm,angle=0}}
\vspace{0.4cm} \centerline{
\epsfig{file=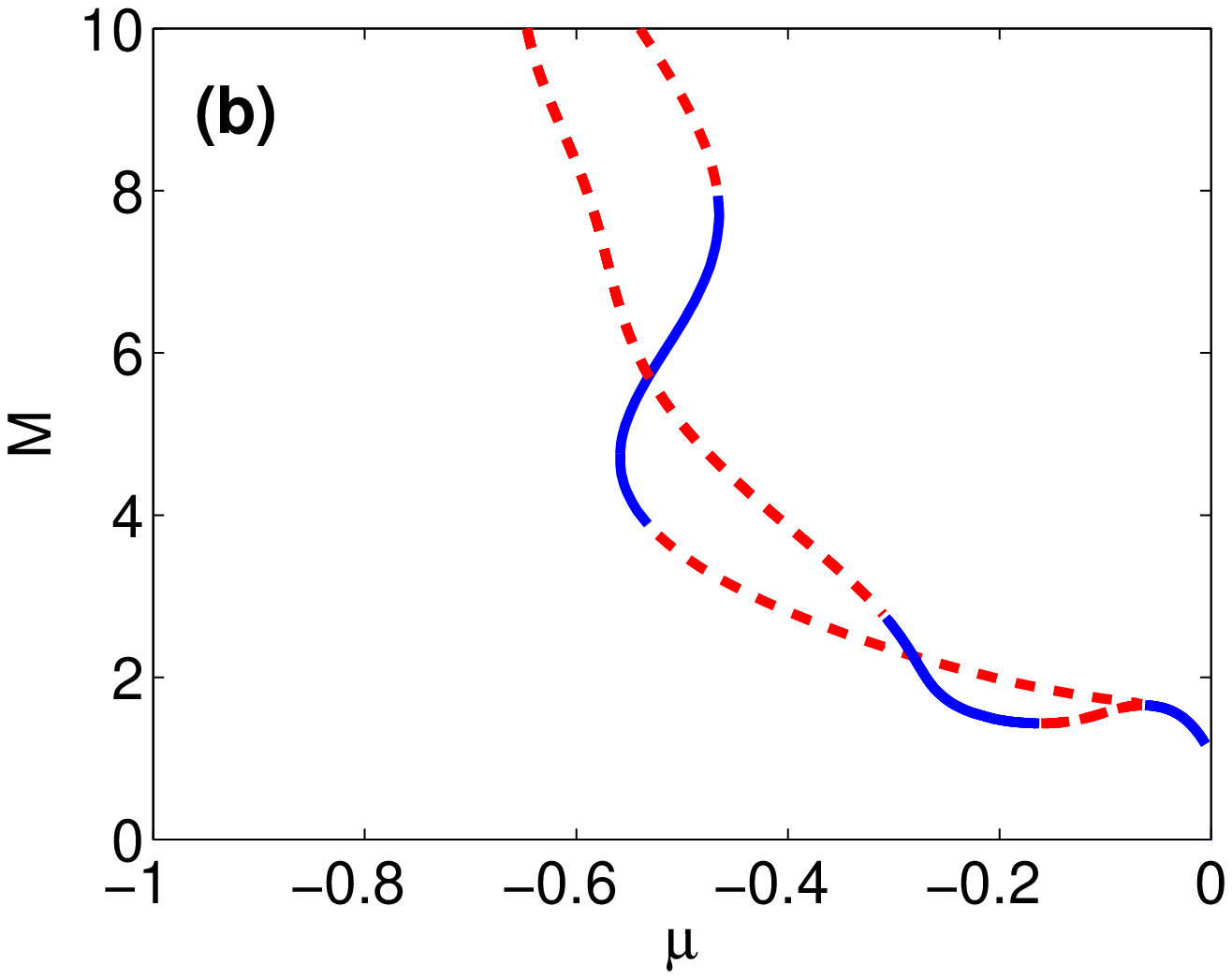,width=6cm,angle=0}
\epsfig{file=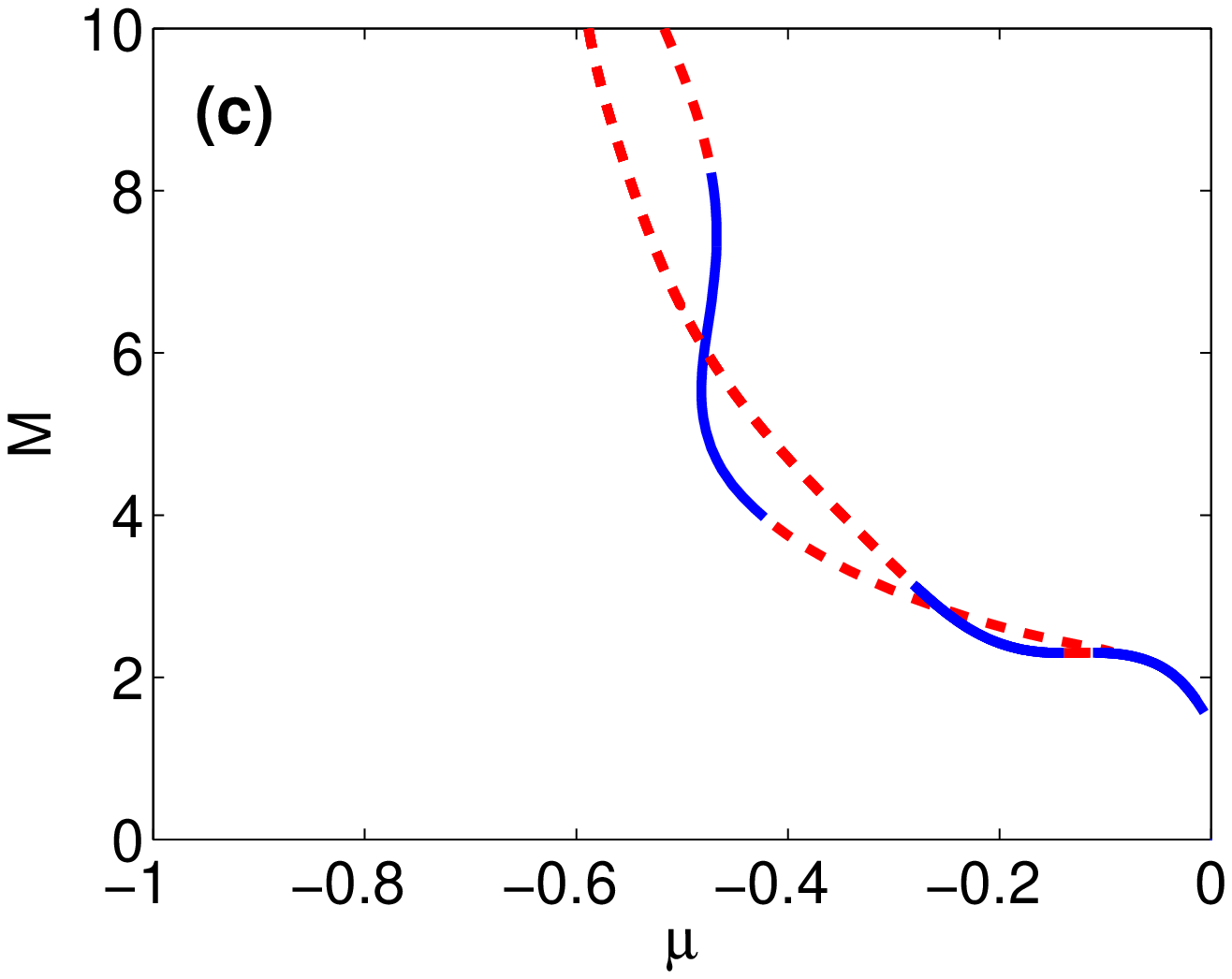,width=6cm,angle=0}
\epsfig{file=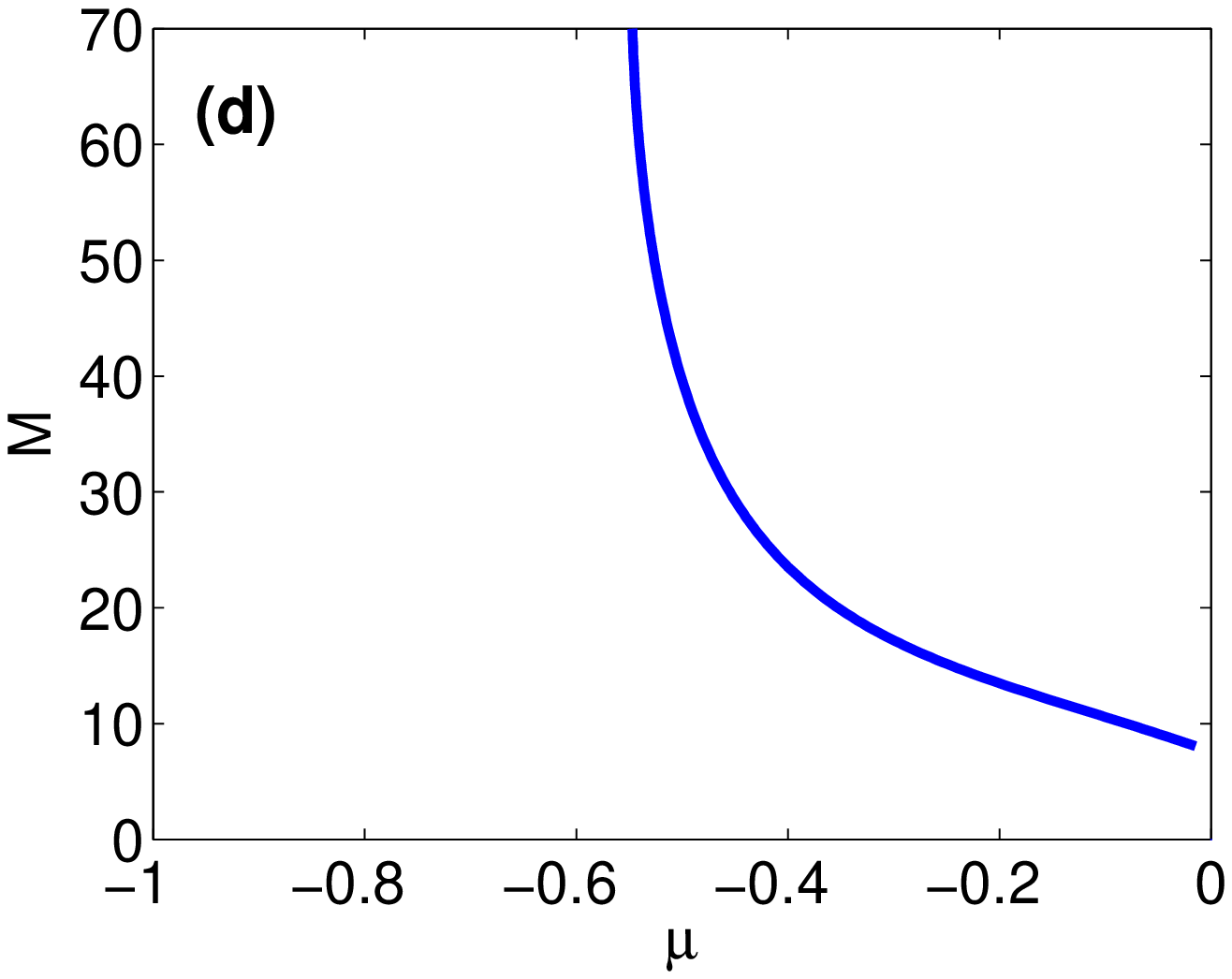,width=6cm,angle=0}}
\caption{(Color online) 
(a) Power ($M$) versus $\mu$ for $C=0.1$, and respective
profiles. Bottom (from left to right): Power diagrams, for (b) $C=0.3$, (c) $%
C=0.4$, and (d) $C=2.0$, of the bond-centered and site-centered
solutions. For low values of $C$ the co-existence of multiple
solutions at different values of $\mu$ is obvious. The
``snaking" pattern gets stretched as $C$ increases, slowly
diminishing the number of solutions until a single solution is left.
Stable and unstable solutions are represented by solid (blue) and
dashed (red) curves, respectively.} \label{snakes.ps}
\end{figure*}

\section{The existence and stability of stationary solutions\label{Sec:bif}}

Detailed existence and stability regions of all above-mentioned solutions
are quite intricate and particularly hard to detect.
%since many solutions types often collide at the same point.
%Indeed, as $C \rightarrow \infty$
%all considered types converge to one single bright profile
%corresponding to the solution in the continuous model.
As described for the 1D case in Ref.~\cite{Ricardo} and mentioned
above, we expect in the 2D case the existence of a large family of
solutions at low values of $C$, which gradually annihilate, through
a series of bifurcations, as $C\rightarrow \infty$
(see Ref.~\cite{konotop} for a detailed description
of the termination scenaria, typically through saddle-node
or pitchfork bifurcations, for the various families
of the basic discrete solitons as the coupling parameter is increased).
By plotting
the power $M$ for various types of the solutions (site-centered,
bond-centered, and hybrid, each with various widths) at fixed
values of $C$ against frequency $\mu $, it is possible to trace the
trend followed by the solutions (see Fig.~\ref{snakes.ps}). For $C=0
$, the exact power for each solution can be found. A snake like
pattern extending from $\mu=-1$ to $\mu=0$ exists and continues
for arbitrarily large powers. This ``snaking"  is also displayed
for different values of $C>0$ in Fig.~\ref{snakes.ps}.  
Branches of the $M(\mu )$ curve with higher powers
correspond to wider solutions. A typical progression observed as one
follows the $M(\mu )$ curve from bottom (low power) to top (high
power) is switching between short and tall solutions with gradually
increasing width. For example, the first branch, which represents
short narrow solutions, collides with a branch of tall narrow
solutions, which then collides with a set of short wide solutions,
and so on. As the coupling strength increases, the power curve gets
stretched upward. Following the stretching, the solutions gradually
vanish, until there remains a single profile. Similar to what was found in
cubic DNLS equation in Ref.~\cite{flach} the bright stationary solutions
in the CQ model also bifurcate from plane waves (near $\mu \approx 0$ for
the CQ model).
It is worthwhile to highlight here the increased 
level of complexity of the relevant $M(\mu )$ curves
in the cubic-quintic model (due to the interplay of short and tall
solution branches) in comparison to its cubic counterpart of Ref.~\cite{pgk2000},
which features a single change of monotonicity (and correspondingly of
stability) between narrow and tall (stable) and wide and short (unstable)
solutions.
%%%%% ENERGY THRESHOLD%%%%%%%%%%

\begin{figure}[tbp]
\centerline{
\epsfig{file=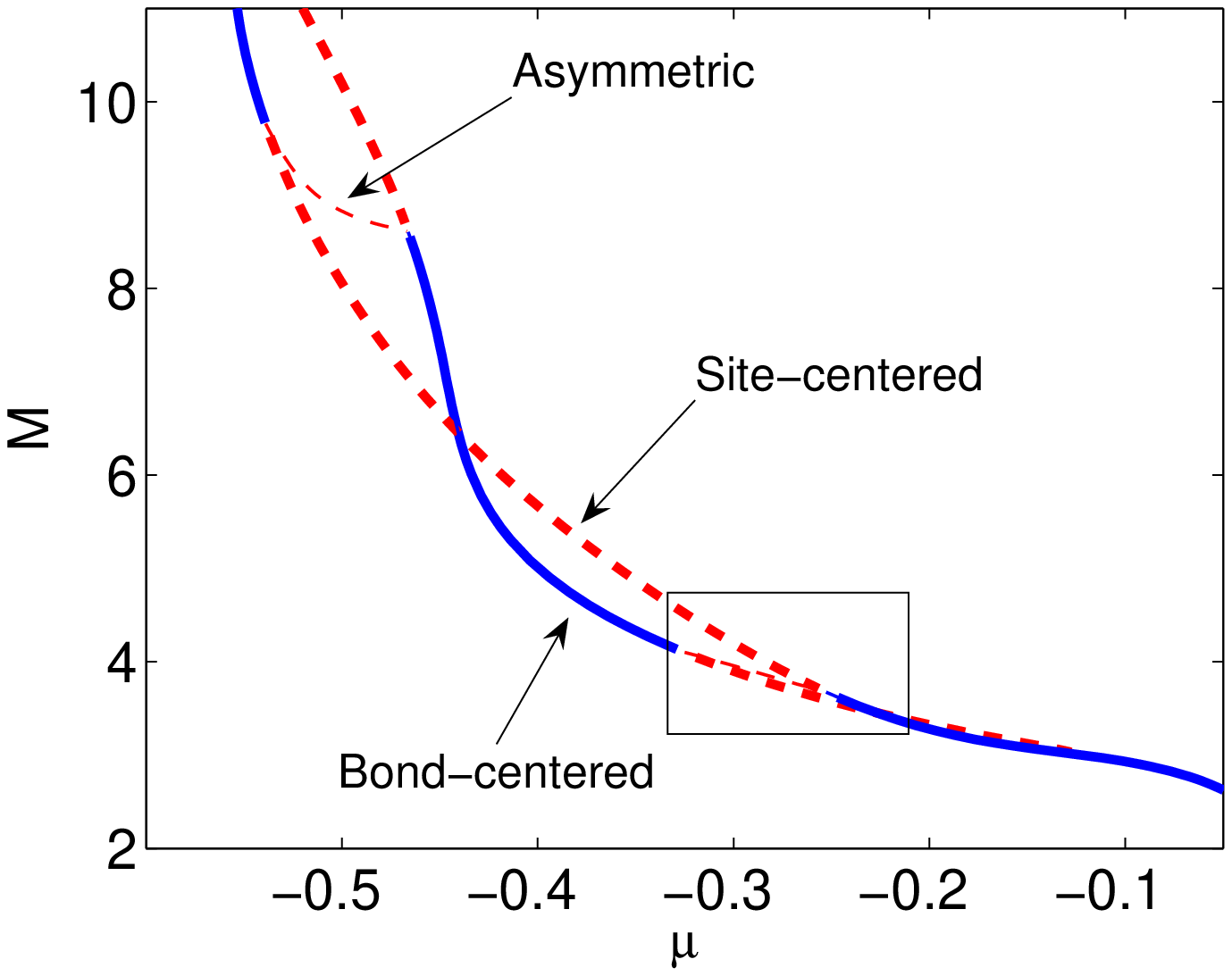,width=7.0cm,angle=0}}
\centerline{
\epsfig{file=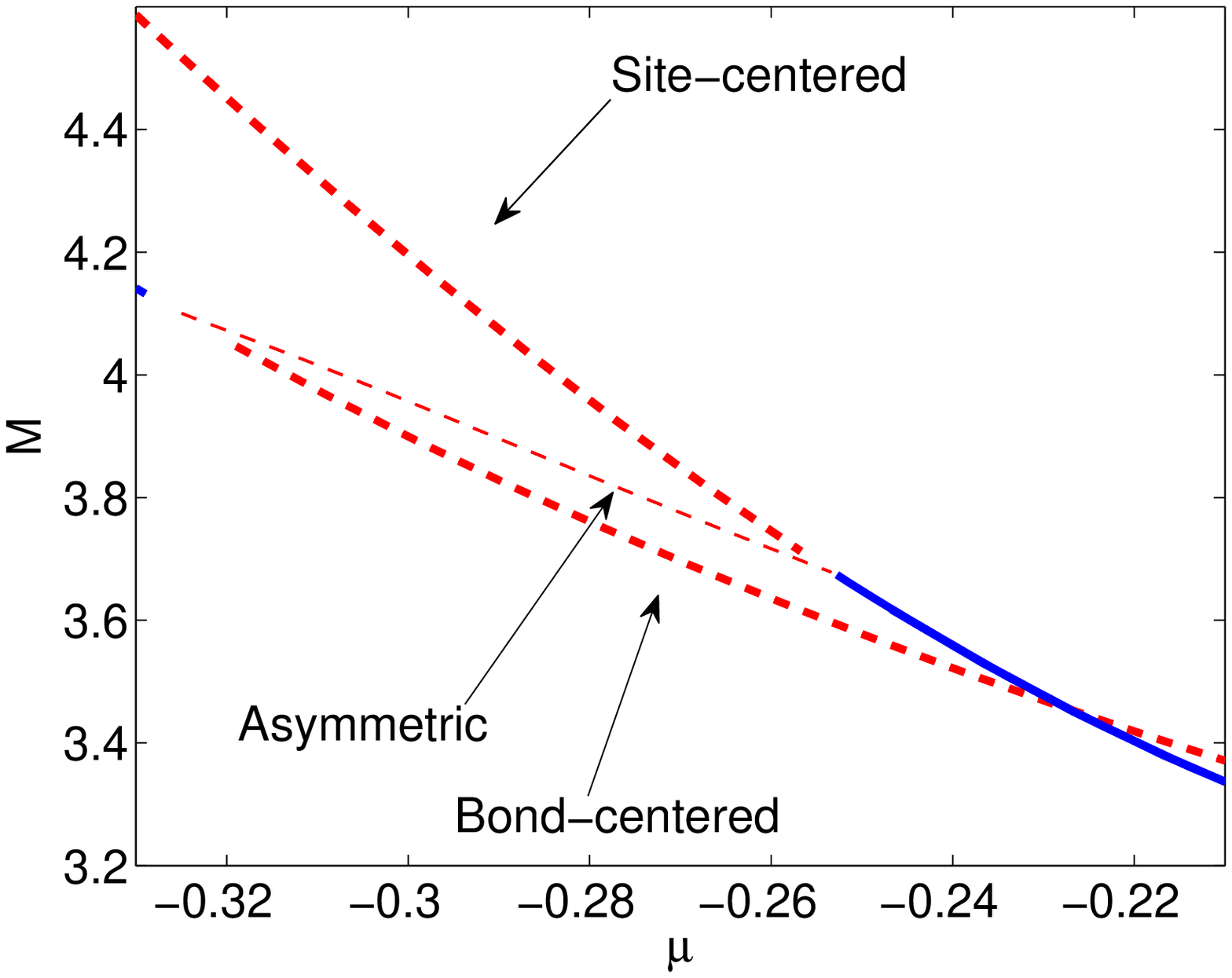,width=7.0cm,angle=0}}
\caption{(Color online) Top: Pitchfork bifurcations of the bond-centered solutions
and site-centered solutions for lattice coupling constant $C=0.5$.
Hybrid solutions are omitted here for clarity. Bottom: Zoom of the
bifurcation scenario depicted by the rectangular region in the
top panel. 
} \label{snake_stab.ps}
\end{figure}

Reference \cite{flach} provides heuristic arguments for the existence of energy thresholds
for a large class of discrete systems with dimension higher
than some critical value. This claim was proved in Ref.~\cite{MIW-2D} for
DNLS models with the nonlinearities of the form $|\psi_n|^{2 \sigma +1} \psi_n$ and 
for chains of NLS equations. 
%We took the opportunity to verify this claim
%in the case of a DNLS with the CQ nonlinearity and found it to be true. 
As can be discerned in Fig.~\ref{snakes.ps}, such thresholds also 
exist in the case of the cubic-quintic nonlinearity.

In Ref.~\cite{Ricardo} a stability diagram for the discrete solitons in the
1D model was presented in the $(\mu ,C)$ plane, which gave a clear overview
of the situation. However, in the present situation, the $M(\mu )$ curves
for various fixed values of $C$, such as those displayed in Fig.~\ref{snakes.ps}, 
provide for a better  understanding of relationships between
different solutions. For example, in the $(\mu ,C)$ diagram, it would appear
that the taller solutions cease to exist at $(\mu ,C)\approx (-0.6,0.4)$.
However, the respective $M(\mu )$ curve shows that narrow and wide solitons
become indistinguishable at this point, and deciding which solution, short or tall,
is annihilated becomes quite arbitrary.

A numerical linear stability analysis was performed in the usual way (see
Ref.~\cite{Ricardo} for details) to investigate the stability of each of the
solution branches. As one follows a $M(\mu )$ curve from bottom to
top, the stability is typically swapped around each turning point, as seen in 
Fig.~\ref{snakes.ps}. However, the stability is not switched exactly at these points,
as this happens via asymmetric solutions (see below).

Similar to the 1D model, a pitchfork-like bifurcation occurs between the
site- and bond-centered discrete solitons. This is more clearly seen in 
Fig.~\ref{snake_stab.ps}. For $C=0.5$, the bond-centered solution loses its
stability in a neighborhood of $\mu \approx -0.53$, and asymmetric solutions
are created there. There are multiple asymmetric
solutions in this case, but only one curve appears in Fig.~\ref{snake_stab.ps}, 
since each one is just a rotation of the other, hence they have
the same power. The bond-centered solution loses its stability
before the site-centered solution regains its
stability; in fact, the site-centered soliton regains the stability exactly
when the asymmetric solutions collide with it. This sort of stability
exchange occurs throughout the $M(\mu )$ curve. The top panel of 
Fig.~\ref{snake_stab.ps} shows two such bifurcations, with a zoom of one of them
shown in the bottom panel.

\begin{figure}[tbp]
\centerline{
\epsfig{file=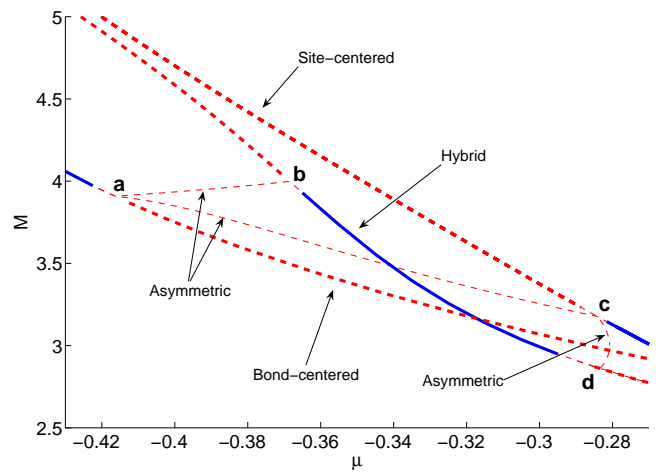,width=10cm,angle=0}}
\caption{(Color online) Bifurcations for $C=0.4$ showing that all three fundamental modes (site-centered,
bond-centered, and hybrid) are all connected to each other  via
stability exchange with asymmetric solutions. Two asymmetric solutions
are created where the bond-centered solution loses stability
at the bifurcation point labeled by `{\bf a}' in the diagram.
One of these asymmetric solutions is connected to the hybrid solution
at `{\bf b}' and the other is connected to the site-centered solution at
point `{\bf c}'.  A third type of asymmetric solution also emanates from
the bifurcation point `{\bf c}' which is connected to the 
hybrid solution at `{\bf d}'. } \label{snake_stab2.ps}
\end{figure}

Figure~\ref{snake_stab2.ps} shows again a site-centered solution
connected to a bond-centered solution but also features a connection of the
site- and bond-centered solutions via the \textit{hybrid} solution.
So not only are all variations (tall, short, narrow, etc) within each mode
connected, as shown by the snake
like power curves, but all the fundamental modes 
(bond-centered, site-centered, hybrid) are also connected.

We stress that the stability regions of the above mentioned fundamental modes
are disjoint in regions
where they each have roughly the same power and, unlike
the 1D model, the asymmetric solutions are \textit{unstable}. These
features can be seen in Figs.~\ref{snake_stab.ps} and \ref{snake_stab2.ps}. Note that the
multi-stability of symmetric solutions still occurs in this case due to the
existence of arbitrarily wide solutions at fixed values of $C$ (see 
Fig.~\ref{snakes.ps}). 
%As explained in the next section, this fact has important
%consequences for the mobility of the lattice solitons.
As a general comment, it should be noted that many of the features of the 2D cubic-quintic
model (such as e.g., the existence of 
unstable asymmetric solutions, and their connecting the fundamental modes)
can also be observed in the case of the saturable model \cite{Sweden},
although in the present case of the cubic-quintic model, the relevant
phenomenology is even richer due to, for instance, the existence of
multiple (i.e., tall and short) steady states.

%%%%%%%%%%%%%%%%%%%%%%%%%%%%%%%%%%%%%%%%%%%%%%%%%%%%%%%%%%%%%%%%%%%%%%%%%%%%%%%
\section{The mobility \label{Sec:mobile}}

In one dimension, traveling solutions can be found in the form
\begin{equation}
\psi _{n}=u(n-vt)e^{i\mu t},  \label{tra}
\end{equation}%
where $v$ is a real velocity. Substitution of this expression in the 1D DNLS
model yields the following advance-delay differential equation
\begin{eqnarray}
0&=& -i[v\dot{u}(z)+i\mu u(z)]+2|u(z)|^{2}u(z)-|u(z)|^{4}u(z)   \notag \\
&&+C\left[ u(z+1)+u(z-1)-2u(z)\right] ,  \label{CQDNLS-tra}
\end{eqnarray}%
where $z=n-vt$. Stationary solutions are said to be \textit{translationally
invariant} if the function $u_{n}=u(nh)$, where $h$ is the lattice spacing, can
be extended to a one-parameter family of continuous solutions, $u(z-s)$, of
the advance-delay equation (\ref{CQDNLS-tra}) with $v=0$. Solutions of this
type have been found in other lattice models (see Ref.~\cite{kev03,Dmitry1} and
references therein). Localized solutions with non-oscillatory tails in
similar models for $v\neq0$, have been found in Ref.~\cite{Melvin,Dmitry2}
by solving a respective counterpart of Eq.~(\ref{CQDNLS-tra}). If
translationally invariant solutions exist, then the sundry modes
(bond-centered, site-centered, etc.) are generated by the same continuous
function $u(z-s)$, each with a corresponding value of $s$. The
translationally invariant solutions occur (i) at \textit{transparency points}%
, which are points in the parameter space where solutions exchange their
stability, and (ii) if the Peierls-Nabarro (PN) barrier vanishes; the
barrier being defined as the difference in energy between the site-centered
and bond-centered solutions. Note that (i) and (ii) are necessary
but not sufficient conditions for the existence of translationally
invariant solutions. For higher-dimensional lattices,
translationally invariant solutions for DNLS-like models have not been found
yet. 
%However, using the notion of translational invariance motives, another
%technique, sometimes referred to as enhanced mobility (explained in detail below), 
%where effectively mobile lattice solitons have been found, even in 2D. They were found in
However, effectively mobile lattice solitons have been found
in 2D models in
regions of the parameter space where the PN barrier is low
(enhanced mobility). This has been the case both 
for quadratic nonlinearities \cite{we} and in the vicinity of
stability exchanges for saturable models \cite{Sweden}.
The resulting mobile solutions radiate energy and eventually come to a halt. 
Exact solutions of the corresponding advance-delay
differential equation, if they exist, would experience no
radiation losses and propagate indefinitely, which is why they are called
radiationless solutions \cite{Melvin}. As mentioned above for translationally 
invariant solutions, radiationless solutions have also
not been yet been found in higher-dimensional lattices. 
In fact, it is an important open question whether such solutions
exist typically, since the single tail resonance appropriately 
made to vanish in Ref.~\cite{Melvin} to obtain such exponentially localized
traveling solutions in 1D settings,
acquires infinite multiplicity in higher dimensional settings. Thus,
the admittedly straightforward technique of identifying
regions of enhanced mobility
may be the only possible method in higher dimensional DNLS problems.

\begin{figure}
\centerline{
\hskip-0.1cm
\epsfig{file=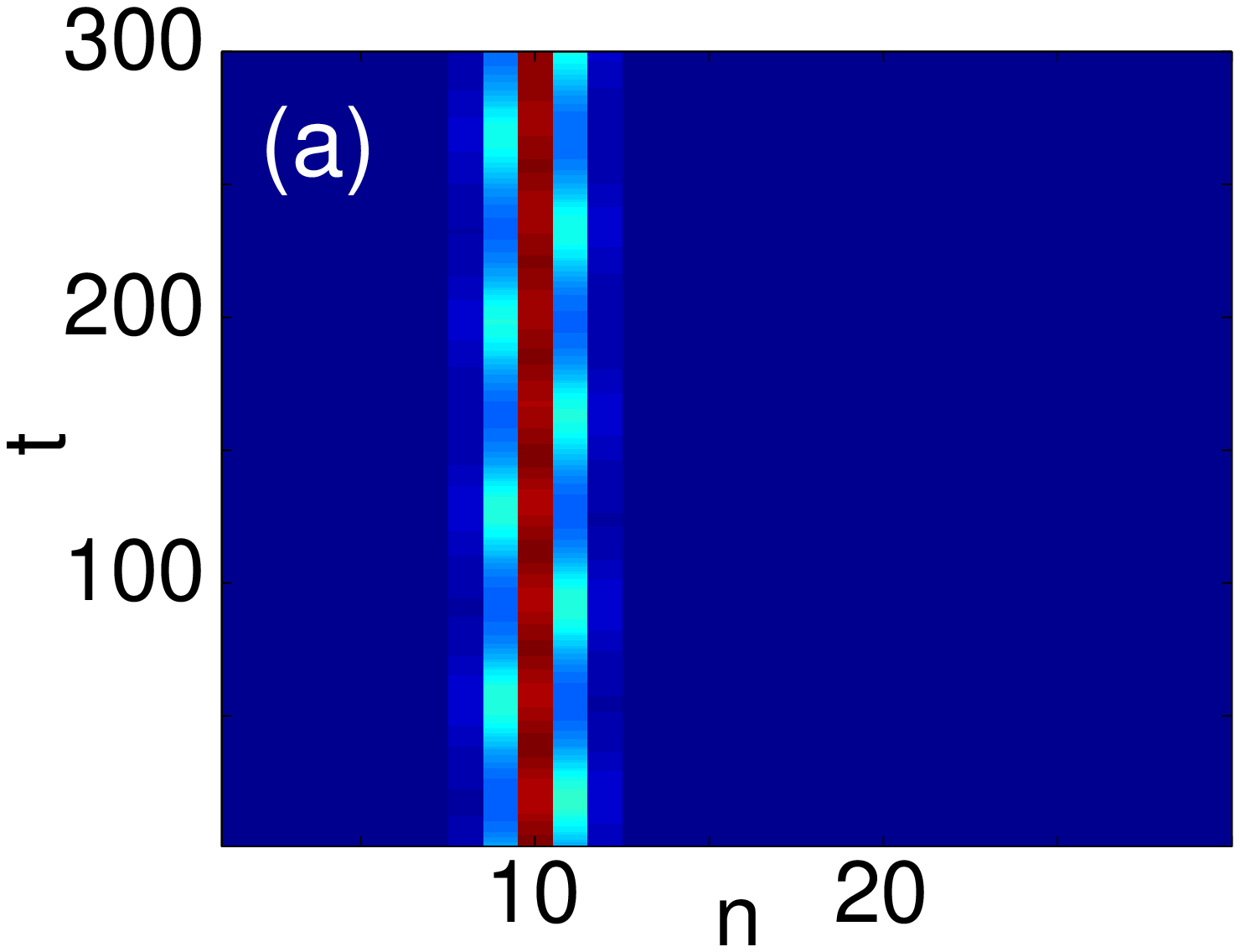,height=2.56cm,angle=0}
\hskip-0.1cm
\epsfig{file=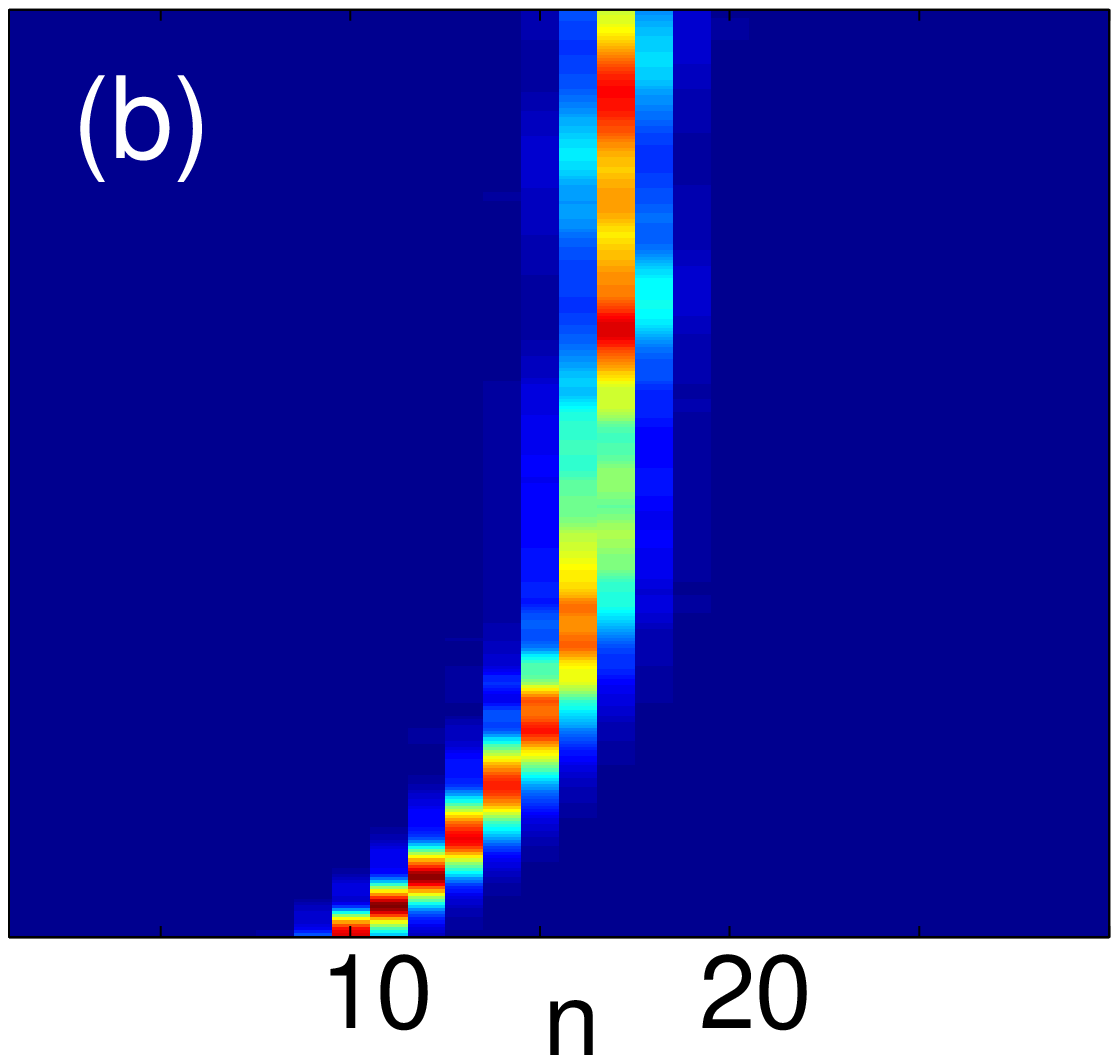,height=2.56cm,angle=0}
\hskip-0.1cm
\epsfig{file=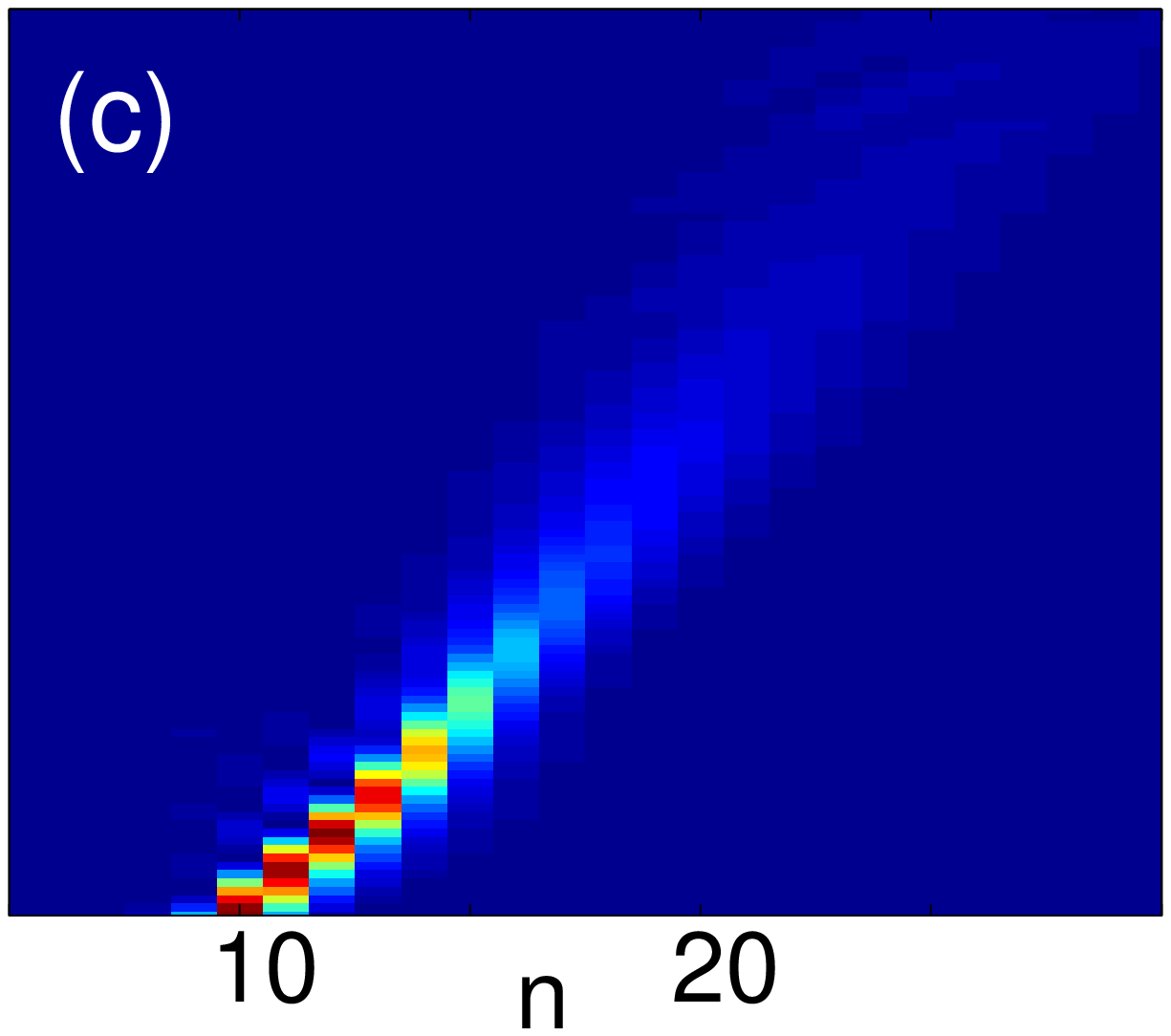,height=2.56cm,angle=0}}
\caption{
(Color online) Resulting density plots, of a one-dimensional
slice along the axis of propagation, from imprinting momentum to a stationary soliton by means of
the ``kick'' defined in Eq.~\ref{kick}.
(a) $k<k_{\mathrm{depin}}$: the solution remains pinned at 
its initial position.
(b) $k_{\mathrm{depin}} < k_n < k_{\mathrm{disperse}}$:
the solution becomes mobile, but eventually comes to a halt due to radiation loss. 
(c) $k>k_{\mathrm{disperse}}$: the kick
is so strong that the solution disperses. }
\label{zones.eps}
\end{figure}

The goal is to ``kick'' the stationary solutions into motion. From a Hamiltonian
point of view, the real part of the solution corresponds to position  and the 
imaginary part to momentum \cite{cubic}. Therefore, in order to set it into 
motion one should apply a perturbation that will alter the imaginary part
of the solution in an asymmetric way, and thus providing it with the necessary momentum
to move. To this end, we apply a ``kick" of the form
\begin{equation}
u_{n,m}(0)=u_{n,m}e^{i(k_{n}n+k_{m}m)},
\label{kick}
\end{equation}
where $u_{n,m}$ is a stationary
solution, and $k_{n}$, and $k_{m}$ are real wavenumbers. This method
has been used in numerous studies in one-dimensional settings
(cf.~Refs.~\cite{cubic,Bang}) and recently in two-dimensions \cite{Sweden}.
Bright mobile solutions were studied in this way in
the 1D CQ model in Ref.~\cite{chongthesis} and
in greater detail (and for staggered solutions) in Ref.~\cite{Belgrade2}.
%
%
%In order to measure the effects of the kick, we define the
%position of the solution along the $n$ axis
%
%\begin{equation} % check this
%<n>(t) = \sum_{n,m} n |\psi_{n,m}|^2 / \sum_{n,m} |\psi_{n,m}|^2.
%\label{pos}
%\end{equation}
%%
%Propagation distance along the  $m$ axis is defined in the same
%way with $m$ replacing $n$ in the above definition. We also
%define a measure of how localized the solution is
%%
%\begin{equation}
%L(t) = \sup_{n,m} |\psi_{n,m}|^2 / \sum_{n,m} |\psi_{n,m}|^2.
%\label{loc}
%\end{equation}
%
We present here results for a site-centered solution 
moving along a single axis only. Therefore we set
$k_n \neq 0$ and $k_m = 0$. Results for other solutions
are similar.
There are three qualitative scenarios that we have observed as result of the kick:
(a) the kick is below some threshold value, $k_n< k_{\mathrm{depin}}$, and so the corresponding 
energy increase is too low to depin the solution, %or more specifically  $<n>(t) \leq \lceil <n>(t_0)\rceil \forall t$  
(b) the kick is greater than this threshold value, $k_n> k_{\mathrm{depin}}$, and the solution is
set in motion eventually halting, or
(c) the initial kick is so strong, $k_n > k_{\mathrm{disperse}}$, that the solution disperses.
% (as indicated by (\ref{loc})).
See Fig.~\ref{zones.eps} for examples of these three scenaria.

\begin{figure}[tbp]
\centerline{
\epsfig{file=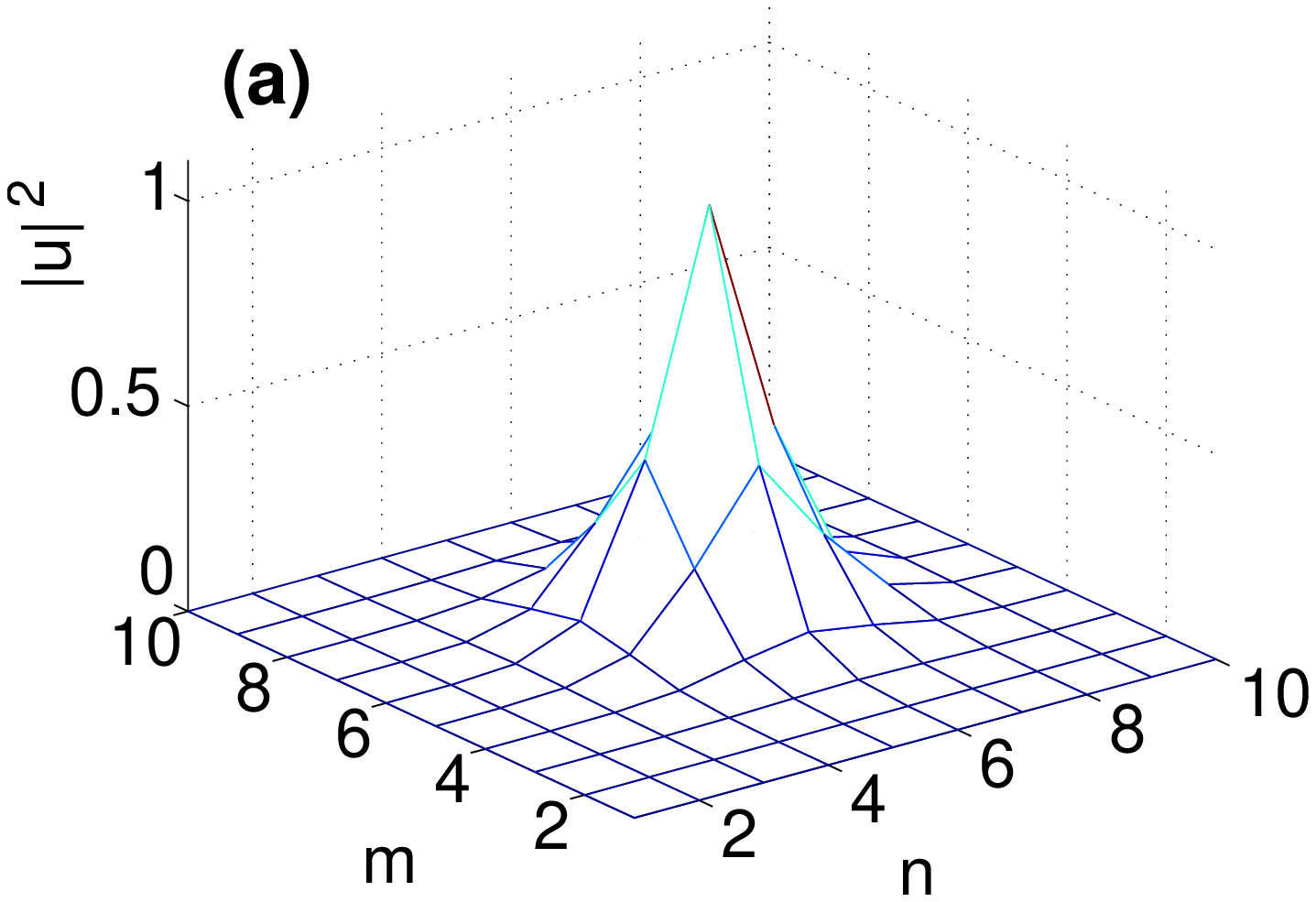,width=4.3cm,angle=0}
\epsfig{file=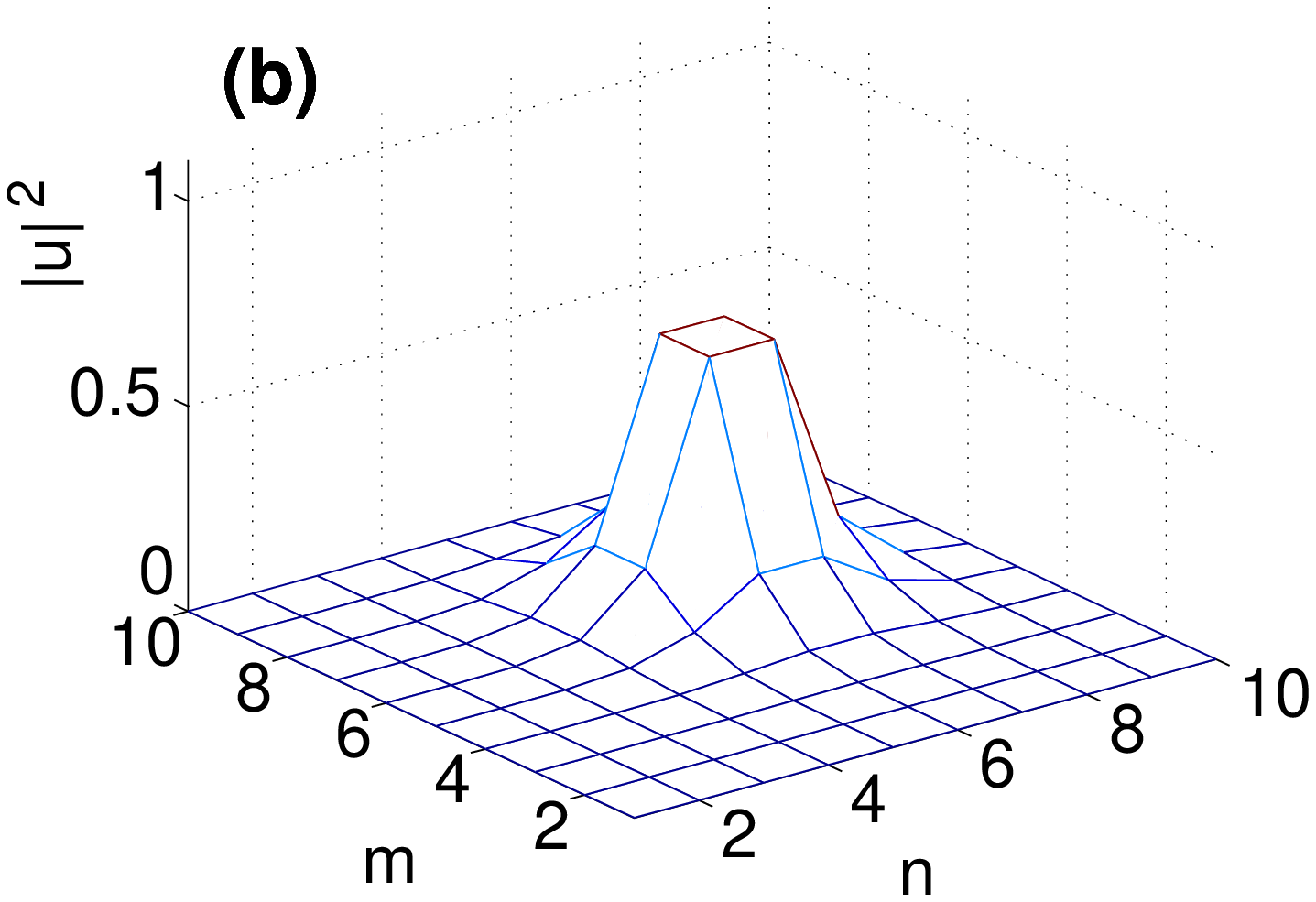,width=4.3cm,angle=0} } \centerline{
\epsfig{file=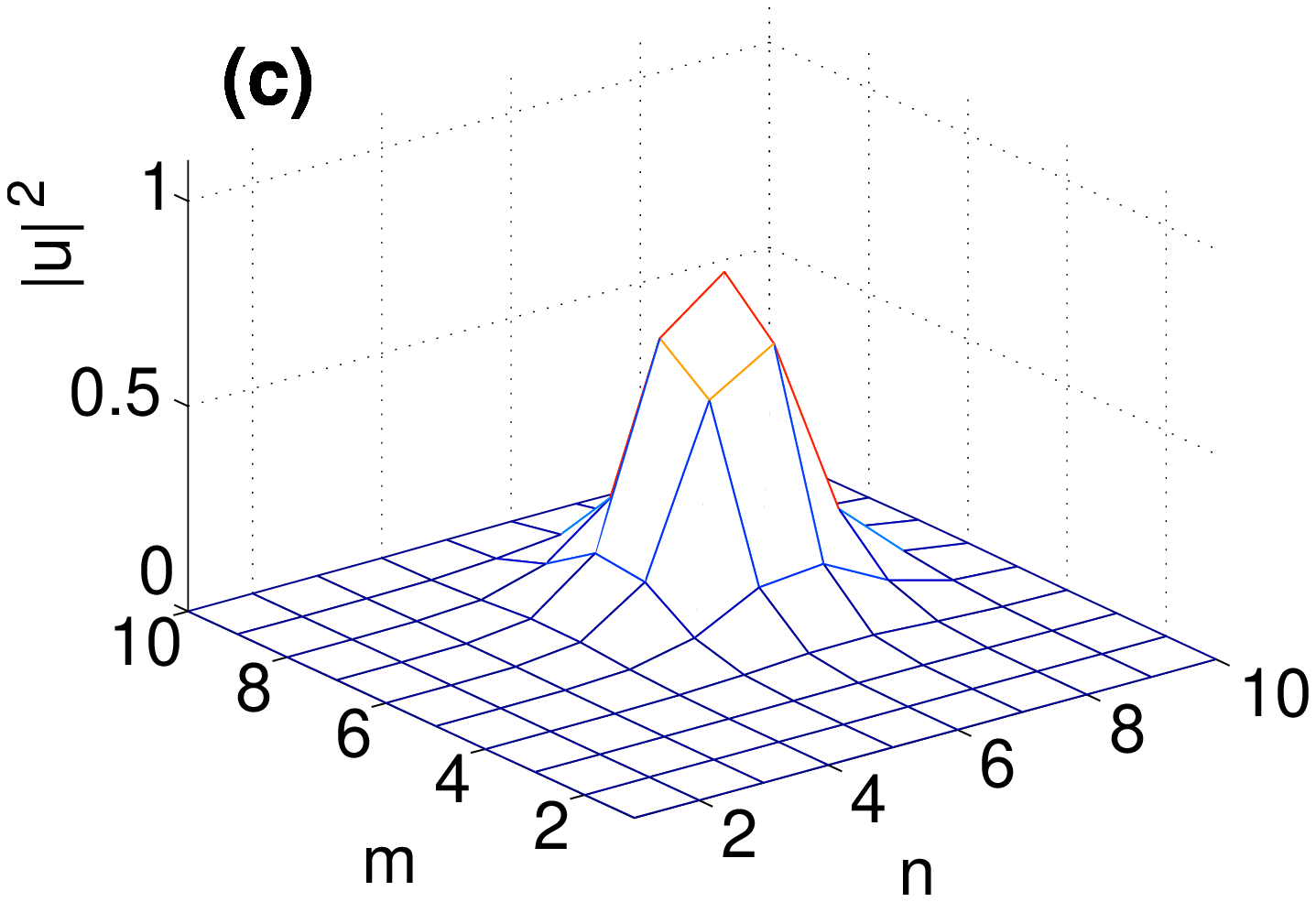,width=4.3cm,angle=0}
\epsfig{file=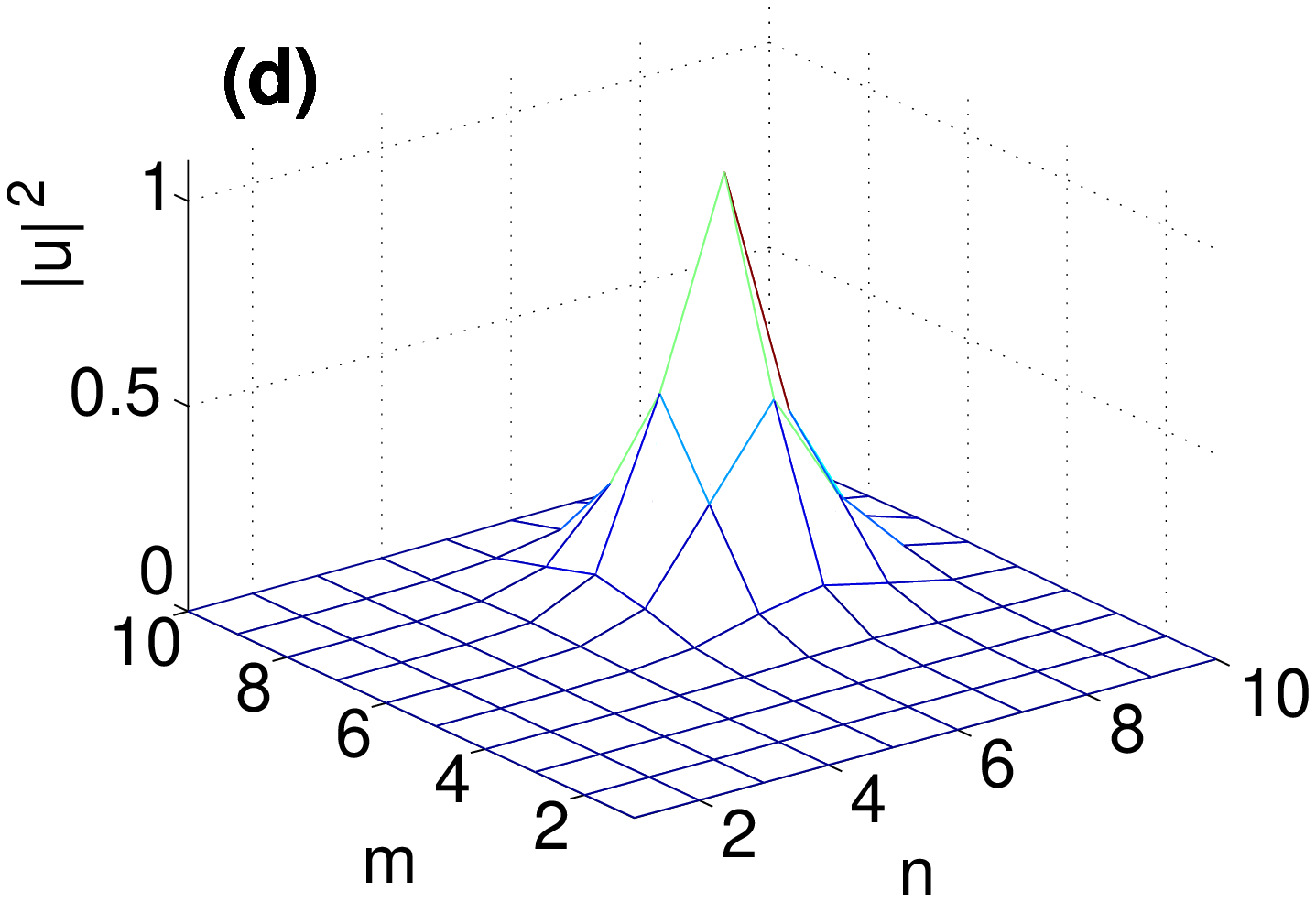,width=4.3cm,angle=0} } 
\caption{(Color online) Evolution
of a site-centered soliton kicked along a diagonal. In the course of
its motion, the traveling object takes on the (a) site-centered,
(b) bond-centered, and (c) asymmetric profiles. This progression
repeats starting again with the (d) site-centered profile until
motion ceases due to radiation loss.} 
\label{deform.ps}
\end{figure}

We are interested in areas of parameter space that provide good conditions
for mobility for the kicked solutions. The PN barrier should provide some insight
as to where these regions may be. While there is no standard
formal definition of the PN barrier in higher
dimensions, one may adopt a natural definition (as used in Ref.~\cite{Sweden}%
), according to which the barrier is the largest energy difference, for a
fixed norm of the soliton, between two stationary solutions of the system
close to configurations that a discrete soliton must pass when moving
adiabatically along the chosen lattice direction. 
This set of configurations includes asymmetric solutions, and,
importantly in higher dimensions, the hybrid solutions too. For example, for a
stationary site-centered soliton to become mobile along an axis, it
must overcome barriers created by the asymmetric and hybrid states,
since, in the course of its motion, its profile will change as follows:
site-centered $\rightarrow $ asymmetric $\rightarrow $ hybrid $%
\rightarrow $ asymmetric $\rightarrow $ site-centered.
If we chose to kick the soliton along the diagonal, then the
same progression should be considered with the bond-centered state replacing the
hybrid one (see Fig.~\ref{deform.ps}). Here we use, for the definition of the
PN barrier,
fixed frequency $\mu$ rather than fixed norm $M$. We also consider the
free energy $G=H-\mu M$ instead of the Hamiltonian as in Ref.~\cite{Melvin}
(note that 
Eq.~(\ref{CQDNLS-c0}) can be derived as $\partial G/\partial \psi_{n,m}^*=0$).

\begin{figure}
\centerline{
\epsfig{file=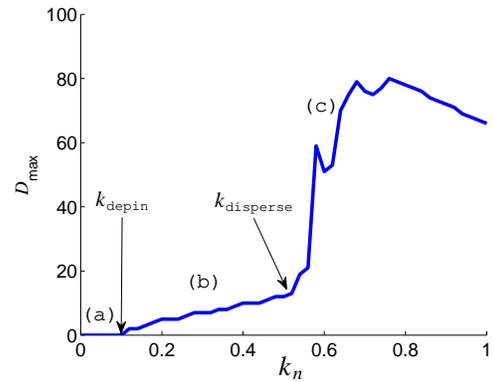,width=7cm,angle=0}}
\caption{The maximum distance traveled as a function of the kicking
strength $k_n$ for $(\mu,C)=(-0.225,0.4)$ and $t \in [0,800]$. 
The area labeled (a) in the graph represents values of $k_n$ that 
could not depin the solution (see Fig.~\ref{zones.eps}.a).
The area labeled by (b) consists of values of $k_n$ that yield a 
mobile solution (see Fig.~\ref{zones.eps}.b) and in (c) the kick is 
so strong that the solution disperses (see Fig.~\ref{zones.eps}.c). 
The threshold values, $k_{\mathrm{depin}}$ and $k_{\mathrm{disperse}}$
are also shown.  }
\label{allzones.eps}
\end{figure}

We kicked the site-centered solutions for various values of 
$k_n$, and estimated the corresponding
threshold values. In Fig.~\ref{allzones.eps} the maximum distance traveled, 
\begin{equation}
D_{\mathrm{max}}(k)= \sup_{t \in [0,T_0]} \lfloor \langle n\rangle(t) \rfloor - \lfloor \langle n\rangle(0) \rfloor, \\
\label{Dmax}
\end{equation}
where the center of mass is computed by
\begin{equation}
\langle n \rangle(t) =  \sum_{n,m} n | \psi_{n,m}(t) |^2 /  \sum_{n,m}  | \psi_{n,m}(t) |^2,  
\end{equation}
is plotted versus the kicking strength. The corresponding threshold values are also
identified there.

\begin{figure*}[tbp]
\epsfig{file=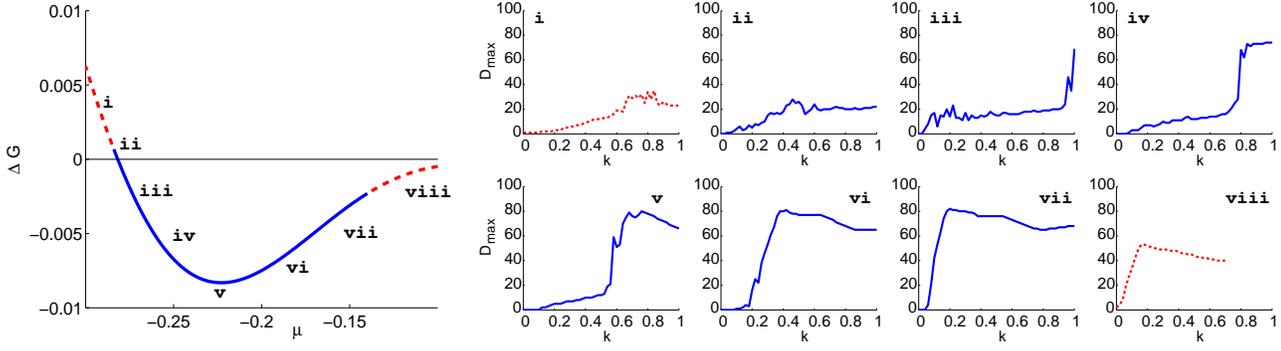,width=17cm,angle=0}
\caption{(Color online) Left: Plot of  $\Delta G_{\mathrm{hybrid}}$ for 
various values of $\mu$ and fixed $C=0.4$. 
The remaining panels (i)--(viii) correspond to the 
maximum distance traveled versus kicking strength plots. See text for
more details.}
\label{Dk.eps}
\end{figure*}

\begin{figure}[th]
\centerline{ 
\epsfig{file=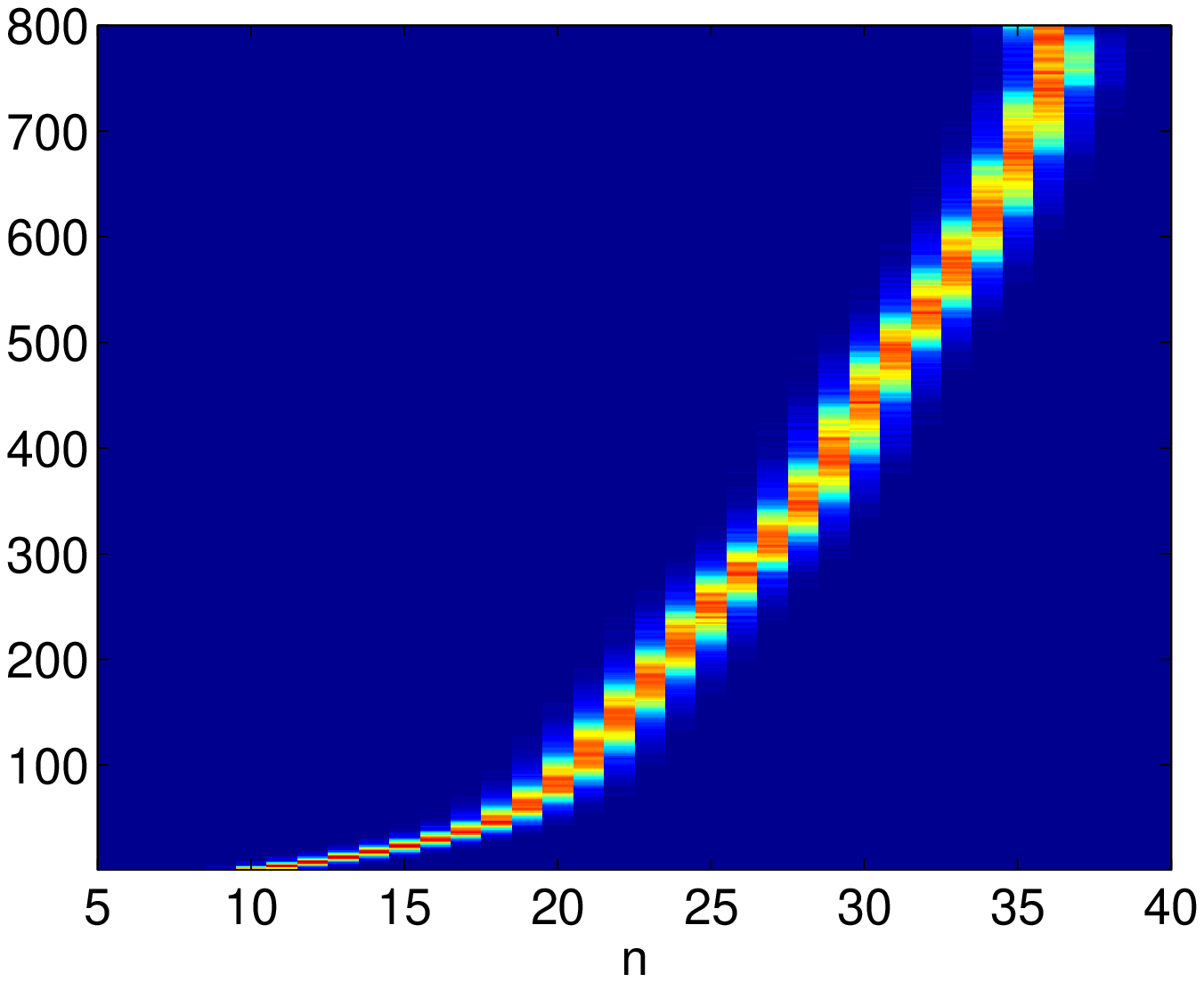,width=6.7cm,angle=0}
}
\centerline{ 
\epsfig{file=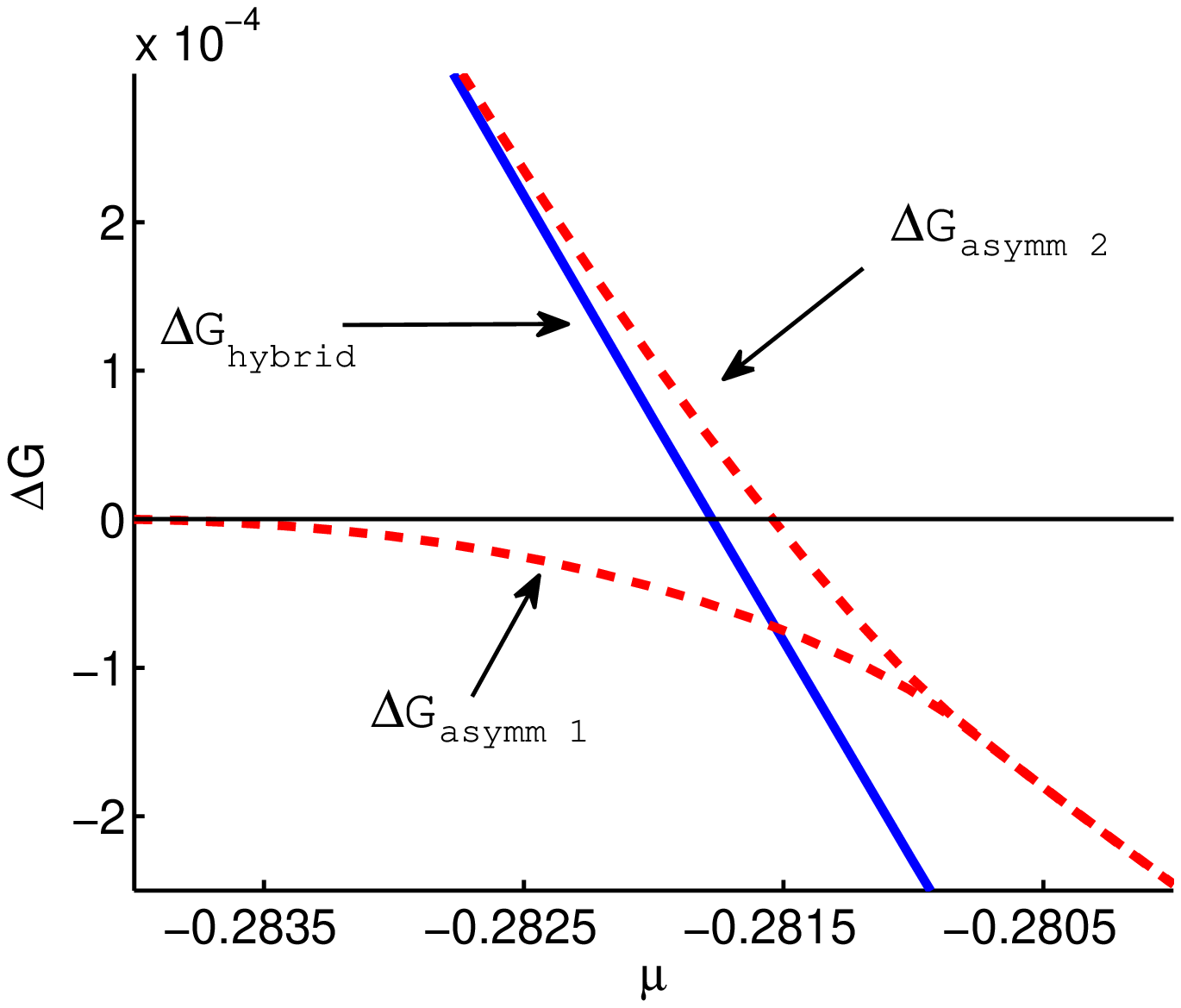,width=6.7cm,angle=0} 
}
\caption{(Color online) 
Top: density plot for the site-centered soliton set in motion along the
lattice axis for $(\mu,C)=(-0.282,0.4)$ and $k_n=0.5$. The choice of parameters
fall in a ``transparency window" 
where good mobility is observed, possibly due to the existence
of a pair of asymmetric solutions.
A one-dimensional slice along the axis of propagation
(at $m=10$) is shown here. Bottom: zoom of the left panel of Fig.~\ref{Dk.eps}
near the ``transparency window". The difference of free energy of the
site-centered solution and the pair of asymmetric solutions 
$\Delta G_{\mathrm{asymm}} = G_{\mathrm{site}} - G_{\mathrm{asymm}}$ is
also shown. The energy added from the kick exceeds
both of these differences. } 
 \label{density.ps}
\end{figure}

It turns out that, the values of the thresholds are related to the
PN barrier. The left panel in Fig.~\ref{Dk.eps} shows the difference
in free energy, $\Delta G_{\mathrm{hybrid}} = G_{\mathrm{site}} - G_{\mathrm{hybrid}}$
between the site-centered solution and the hybrid solution for
fixed $C=0.4$ and $\mu \in [-0.3,-0.1]$. In each subpanel of the figure
$D_{\mathrm{max}}$, as defined in Eq.~(\ref{Dmax}), is plotted
against the kicking strength for $t \in [0,800]$ for fixed $\mu$.
In panel (i) the site-centered solution has more energy
than the hybrid solution but is unstable and moves away from its initial position even for $k_n=0$.
Panel (ii) represents parameter values where the site-centered solution
has greater power and is stable. In this small ``transparency window" of parameter space,
there is also pair of unstable asymmetric solutions. In this region,
we observed the best mobility (see Fig.~\ref{density.ps}). This
is consistent with what was found in the saturable 2D DNLS \cite{Sweden}
where good mobility was found where asymmetric solutions exist. In panel (iii) the threshold
$k_{\mathrm{disperse}}$ is visible and the sign of $\Delta G_{\mathrm{hybrid}}$ has switched. In (iv)
we see that the value of $k_{\mathrm{depin}}$ is increasing and 
$k_{\mathrm{disperse}}$ is decreasing as the PN barrier increases. 
Panel (v) corresponds to the maximum
energy difference. This is also where the largest $k_{\mathrm{depin}}$
occurs. As the energy difference decreases once again as seen in (vi)
the threshold $k_{\mathrm{disperse}}$ continues to decrease. This is also
the case in panel (vii) as both thresholds approach $k_n=0$. Finally, for the
unstable region in (viii) $k_{\mathrm{depin}}$ is once again zero.

We were unable to identify true transparency points in the present model
(the 2D DNLS lattice with the CQ onsite nonlinearity), which 
seems to preclude the
possibility of finding exact translationally invariant solutions. However,
enhanced mobility was achieved by lending stationary solutions kinetic
energy in cases where the PN barrier was low. These moving states gradually
lose energy and get eventually trapped at some positions in the lattice.
Solving higher-dimensional counterparts of Eq.~(\ref{CQDNLS-tra}) in the
higher-dimensional lattice might reveal moving radiationless solutions, 
although, as we pointed out above, solutions to this (quite difficult) %numerical 
problem may not typically exist.
%but
%a solution of this (quite difficult) problem is beyond the scope of the
%present work.
%
It is worth mentioning in passing that the energy loss in the 1D discrete 
sine-Gordon lattice has been recently described using an 
averaged Lagrangian approach in Ref.~\cite{minzoni:08}.

\section{Three-dimensional solutions \label{Sec:3D}}

\begin{figure}[th]
%\centerline{
% \epsfig{file=\rootfig site3Dprofile.eps.jpg.ps,width=4.5cm,angle=0}
% \epsfig{file=\rootfig bond3Dprofile.eps.jpg.ps,width=4.5cm,angle=0}
% }
%\centerline{
% \epsfig{file=\rootfig hybrid13Dprofile.eps.jpg.ps,width=4.5cm,angle=0}
% \epsfig{file=\rootfig hybrid23Dprofile.eps.jpg.ps,width=4.5cm,angle=0}
%  }
\centerline{
\epsfig{file=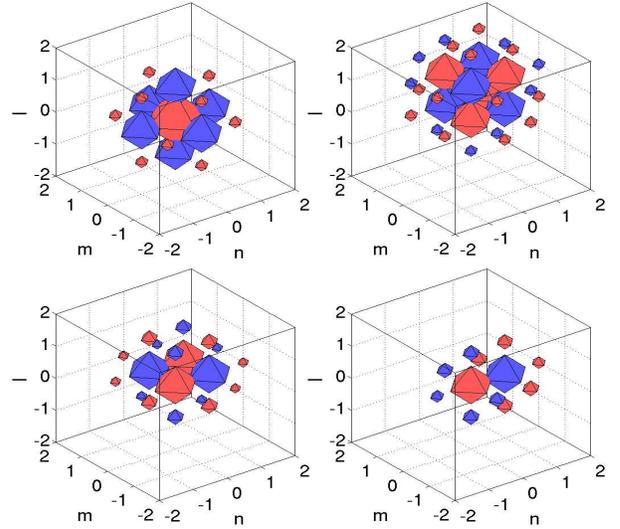,width=8.0cm,height=7cm,angle=0}
}
\caption{(Color online) 
Plot of the basic configurations in the 3D lattices using iso-contours.
Top: plot of 3D site-centered (left) and bond-centered
(right) solitons. Larger diamonds correspond to larger local amplitudes.
Bottom: Two different types of 3D hybrid solutions. 
The different colors (arranged in a 3D check-board pattern)
are solely used for clarity of presentation.}
\label{3Dprofiles.ps}
\end{figure}

We will now briefly consider a 3D version of the CQ DNLS model. The
respective counterpart of Eq.~(\ref{CQDNLS-sta}) is
\begin{eqnarray}
i\dot{\psi}_{n,m,l} &+&C\Delta ^{(3)}\psi _{n,m,l}+2|\psi _{n,m,l}|^{2}\psi
_{n,m,l}  \notag \\[1ex]
&-&|\psi _{n,m,l}|^{4}\psi _{n,m,l}=0,  \label{3DCQDNLS}
\end{eqnarray}%
where $\psi _{n,m,l}$ is the complex field at site \{$n,m,l$\}. In an
isotropic medium, the discrete Laplacian is
\begin{eqnarray}
\Delta ^{(3)}\psi _{n,m,l} &\equiv& \psi _{n+1,m,l}+\psi
_{n-1,m,l}+\psi _{n,m+1,l}+\psi _{n,m-1,l}\nonumber
 \\[1ex]
&&+\psi _{n,m,l+1}+\psi _{n,m,l-1}-6\psi _{n,m,l}.
\end{eqnarray}%
We search for stationary solutions, $\psi _{n,m,l}=u_{n,m,l}\exp (-i\mu t)$,
using the same method as in Sec.~\ref{Sec:model}. The 2D soliton species
have their natural 3D counterparts. As shown in Fig.~\ref{3Dprofiles.ps},
the extra dimension admits an additional type of a hybrid soliton.

\begin{figure}[tbp]
\centerline{
 \epsfig{file=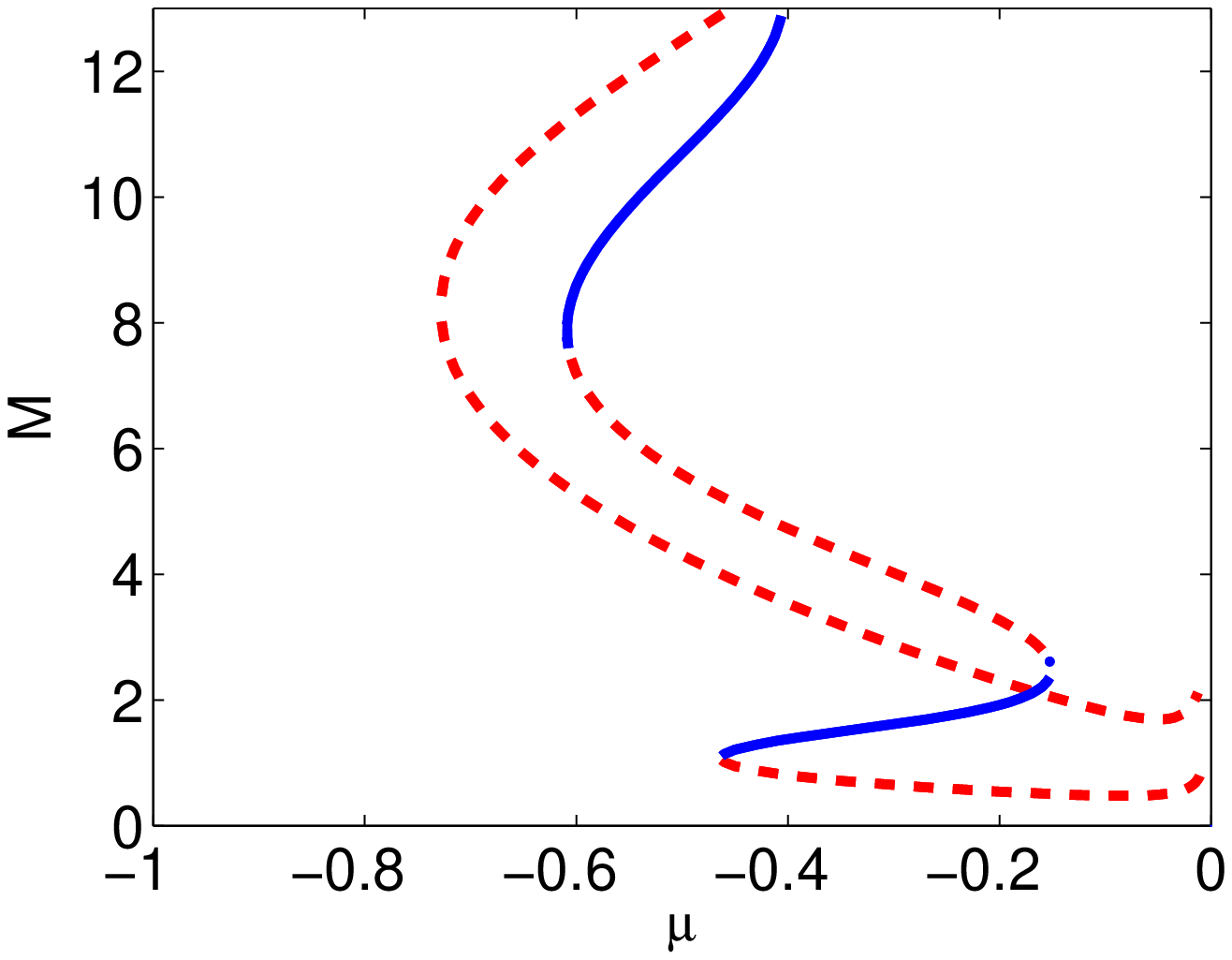,width=7.0cm,angle=0}
 }
\centerline{
 \epsfig{file=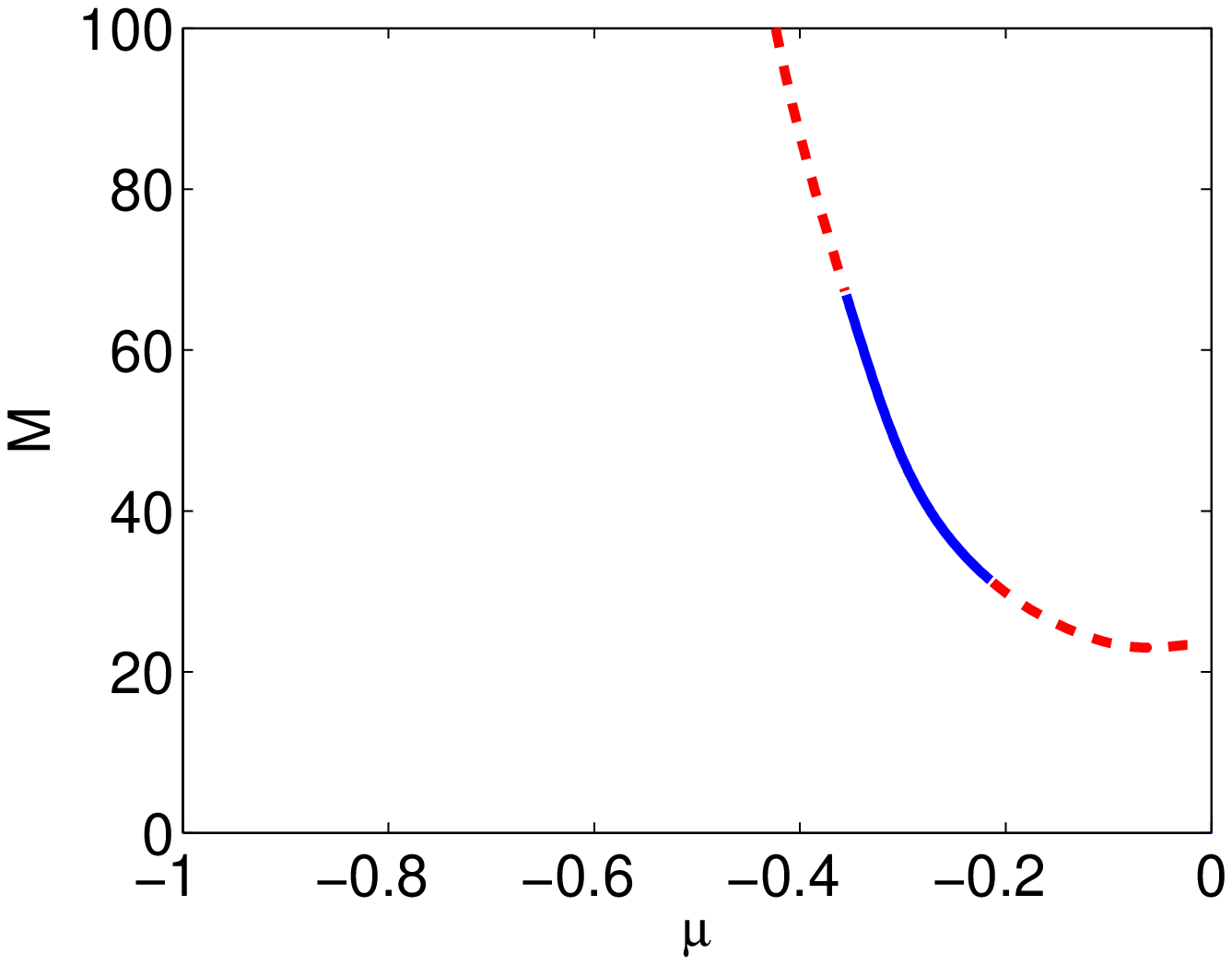,width=7.0cm,angle=0}
 }
\caption{(Color online) The power of the site- and bond-centered solitons versus the
frequency for $C=0.1$ (top) and $C=0.7$ (bottom) in the 3D lattice.}
\label{3Dbif.ps}
\end{figure}

Figure \ref{3Dbif.ps} shows $M(\mu )$ curves for 3D bond-centered and
site-centered solitons for $C=0.1$ and $C=0.7$. The figure illustrates
that in the 3D case, 
similarly to the 2D case, the
snake-like patterns are present for small values of coupling constant $C$
and are stretched as $C$ is increased. Similar results were obtained for
the 3D hybrid solutions (results not shown here).

%%%%%%%%%%%%%%%%%%%%%%%%%%%%%%%%%%%%%%%%%%%%%%%%%%%%%%%%%%%%%%%%%%%%%%%%%%%%%%%%%%%%%%

\section{Variational approximation \label{Sec:VA}}

Following the pattern of the VA developed in Ref.~\cite{Ricardo} for 1D
discrete solitons in the CQ-DNLS model, it is possible to construct
analytical approximations for the discrete solitons, and compare them to the
numerical solutions described above. We present this approach for the 2D
model, but the procedure is essentially the same in three dimensions. It is
relevant to mention that the VA for 1D discrete solitons in models of the
DNLS type was first developed in Ref.~\cite{MIW-1D}.
%, and then it was
%generalized for 2D models in Ref.~\cite{MIW-2D}.

Solutions to the stationary version Eq.~(\ref{2DCQDNLS}) are local extrema
of the corresponding Lagrangian,
\begin{eqnarray}
{L} &=&\displaystyle\sum_{n,m=-\infty }^{\infty }\mu u_{n,m}^{2}+u_{n,m}^{4}-%
\frac{1}{3}u_{n,m}^{6}  \label{Lag2} \\[1ex]
&-&C\left[ (u_{n+1,m}-u_{n,m})^{2}+(u_{n,m+1}-u_{n,m})^{2}\right]  \notag
\end{eqnarray}%
[recall $\psi _{n,m}=u_{n,m}\exp (-i\mu t)$]. We approximate each soliton by
a localized ansatz which makes it possible to evaluate the infinite sums in 
Eq.~(\ref{Lag2}) in an explicit form. First, the following ansatz is used for the
site-centered (sc) solution:

\begin{equation}
u_{m,n}^{\mathrm{(sc)}}=\left\{
\begin{array}{ll}
\beta  & \mathrm{~if~} m=n=0, \\[1ex]
Ae^{-\alpha (|m|+|n|)} & \mathrm{~otherwise} \end{array}%
\right.  \label{ansatzST}
\end{equation}%
where $A,\beta ,$ and $\alpha $ are real constants to be found from the
Euler-Lagrange equations,

\begin{equation}
\frac{\partial L_{\text{eff}}}{\partial A}=\frac{\partial L_{\text{eff}}}{%
\partial \alpha }=\frac{\partial L_{\text{eff}}}{\partial \beta }=0,
\label{Euler-Lagrange3}
\end{equation}%
$L_{\text{eff}}$ standing for Lagrangian (\ref{Lag2}) evaluated with ansatz (%
\ref{ansatzST}). In particular, $\alpha $ is treated here as one of
the variational parameters, in contrast to the 1D case, where it was
expressed in terms of  $\mu$ and $C$ by means of a relation obtained
from the consideration of the linearized stationary equation for
decaying ``tails" of the soliton \cite{Ricardo},
\begin{equation}
\alpha =\ln \left( \frac{a}{2}+\sqrt{\left( \frac{a}{2}\right) ^{2}-1}%
\right) ,~a\equiv 2-\mu /C.  \label{eigen}
\end{equation}%
We have observed, based on numerous calculations, that treating $\alpha$ 
as a variational parameter yields the same
relation for $\alpha$ in both the 2D and 3D models. This
is consistent with solutions in the continuum model where 
it is known that the factor in the exponential tail 
is independent of the dimension\footnote{In the continuum model the
tail decays as $r^{-1/2}e^{-br}$ in the 2D case and as 
$r^{-1}e^{-br}$ in the 3D case where the factor $b$
is independent of the dimension.}.

\begin{figure}[tbp]
\centerline{
 \epsfig{file=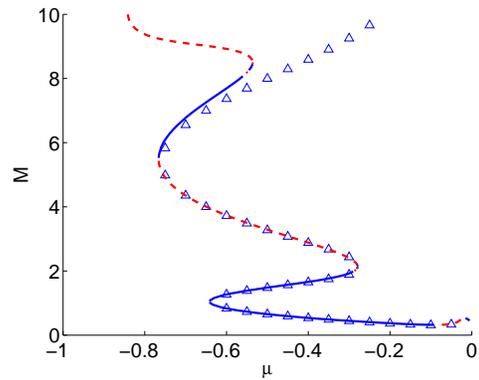,width=7.0cm,angle=0}
 }
\caption{Numerical solutions (solid line) and the variational approximation
(triangles) for the site-centered solitons at $C=0.1$ %(top) and $C=0.5$ (bottom)
in the 2D lattice model. The approximation based on ansatz (\ref{ansatzST}) 
is able to capture subfamilies of tall and short narrow
solitons, and the branch of short wide solitons too.}
\label{2DvaC1.ps}
\end{figure}

%Solving Eqs.~(\cite{Euler-Lagrange1})--(\cite{Euler-Lagrange3}) gives
%us values for the parameter of the approximate solution given by
%the ansatz (\ref{ansatzST}) which does well for low
%values of the coupling constant $C$ (see Fig.~\ref{2DvaC1.ps}).
Solutions predicted by the VA based on ansatz (\ref{ansatzST}) provide for a
good fit to the short and tall narrow solutions and the first subfamily of
wide short solitons of the site-centered type, (see Fig.~\ref{2DvaC1.ps}). At
larger values of $C$, the VA-predicted solutions depart from the numerical
ones, which is not surprising, as the exponential cusp implied by the ansatz
is not featured by the discrete solitons in the strong-coupling
(quasi-continuum) model.

Other solution types can be approximated by 
appropriately modified \textit{ans\"{a}tze}.
In particular, the bond-centered (bc) soliton is based on a frame built of
four points with equal amplitudes (see Fig.~\ref{2Dprofiles.ps}.b), 
whereas the hybrid (hy) soliton has just
two points in its frame (see Fig.~\ref{2Dprofiles.ps}.c). 
Accordingly, the solitons of these types can be
modeled by the following modifications of ansatz (\ref{ansatzST}):
\begin{equation}
u_{m,n}^{\mathrm{(bc)}}=\left\{
\begin{array}{ll}
\beta                 &  ~m,n \in \{0,1\} \\[1ex]
Ae^{-\alpha (|m|+|n|)} & \mathrm{~if~}m,n<0 \\[1ex]
Ae^{-\alpha (|m-1|+|n|)} & \mathrm{~if~}m > 1, n < 0 \\[1ex]
Ae^{-\alpha (|m|+|n-1|)} & \mathrm{~if~}m < 0, n > 1 \\[1ex]
Ae^{-\alpha (|m-1|+|n-1|)} & \mathrm{~otherwise}
\end{array}%
\right.  \label{ansatzPage}
\end{equation}%
and
\begin{equation}
u_{m,n}^{\mathrm{(hy)}}=\left\{
\begin{array}{ll}
\beta                 & ~n=0, m \in \{0,1\} \\[0.5ex]
Ae^{-\alpha (|m|+|n|)} & \mathrm{~if~}m,|n|<0 \\[0.5ex]
Ae^{-\alpha (|m-1|+|n|)} & \mathrm{~otherwise}
\end{array}%
\right.  \label{ansatzHybrid}
\end{equation}

\begin{figure}[tbp]
\centerline{
 \epsfig{file=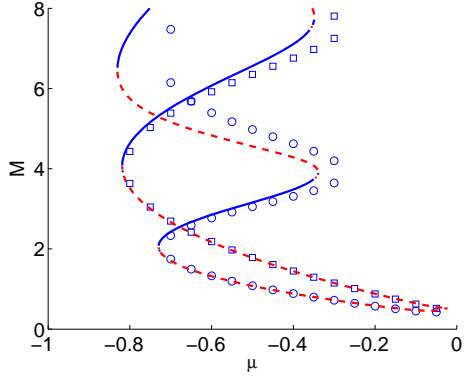,width=7.0cm,angle=0}
 }
\caption{Numerical solutions (solid line) and the variational approximation
for the bond-centered (squares) and hybrid (circles) solitons at $C=0.1$
in the 2D lattice model. The approximations based on the \textit{ans\"{a}tze}
given in (\ref{ansatzPage}) and (\ref{ansatzHybrid}) respectively
are able to capture subfamilies of tall and short narrow
solitons.}
\label{2DvaC1_bond_hy.ps}
\end{figure}

Further analysis demonstrates that the
modified \textit{ans\"{a}tze} produce a good approximation for the
short and tall narrow solutions at small $C$ but not any of the wide
families (see Fig.~\ref{2DvaC1_bond_hy.ps}).

\begin{figure}[tbp]
\centerline{
 \epsfig{file=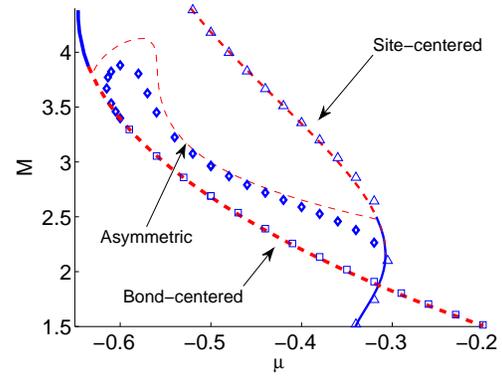,width=7.0cm,angle=0}
 }
\caption{Bifurcations featuring the bond-centered, site-centered, and
asymmetric solutions for $C=0.22$. Numerical solutions (lines) and 
its predicted counterparts using the VA based on 
the ansatz (\ref{ansatzAsym}) 
(markers) are in good agreement. The asymmetric
VA solution captures the main qualitative features of the $M(\mu)$ curve 
(e.g. the dramatic increase of power around $\mu \approx -0.55$) but
slightly underestimates the power at the bifurcation points. }
\label{vabif.eps}
\end{figure}

We were also able to predict complicated bifurcations of the system
by introducing the appropriately chosen asymmetric (asym) ansatz:
\begin{equation}
u_{m,n}^{\mathrm{(asym)}}=\left\{
\begin{array}{ll}
\beta_1                 & ~n=0, m =0 \\[0.5ex]
\beta_2                 & ~n=0, m =1 \\[0.5ex]
\beta_3                 & ~n=1, m =0 \\[0.5ex]
\beta_4                 & ~n=1, m =1 \\[0.5ex]
Ae^{-\alpha (|m-\zeta|+|n-\zeta|)} & \mathrm{~otherwise}
\end{array}%
\right.  \label{ansatzAsym}
\end{equation}
The intention here is to capture the bifurcations where
the site-centered and bond-centered solutions are
connected via an asymmetric solution. Therefore 
we have some idea \textit{a priori} what the asymmetric
solutions should look like and have chosen
ansatz (\ref{ansatzAsym}) accordingly. For $\zeta=0$
the ansatz has the form of a site-centered solution whereas
for $\zeta=0.5$ it will represent a bond-centered solution.
All intermediate values of $\zeta$ represent
asymmetric solutions that are somewhere between
a site-centered and bond-centered solution. Indeed,
the computed value of $\zeta$ based on the variational
approximation starts near $\zeta=0.5$ for parameter values where the
asymmetric solution is almost connected to bond-centered solution,
and slowly decreases to $\zeta=0$ as we alter the parameters
until it collides with the site-centered solution
(see Fig.~\ref{vabif.eps}).

\begin{figure}[tbp]
\centerline{
 \epsfig{file=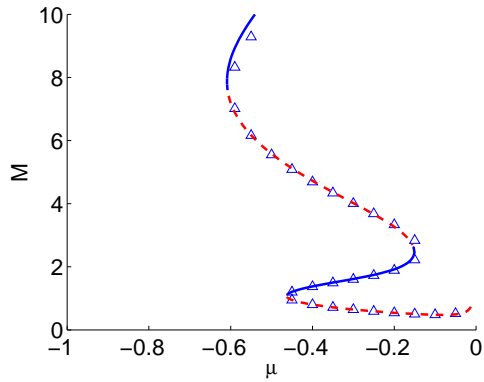,width=7.0cm,angle=0}
 }
\caption{Numerical solution (solid line) and the variational approximation
(triangles) for the site-centered solitons with $C=0.1$ using
the ansatz given in (\ref{ansatzST3D}) %(top) and $C=0.5$ (bottom)
in the 3D DNLS lattice with cubic-quintic nonlinearities.}
\label{3Dva1.ps}
\end{figure}

%\textcolor{red}{[\textrm{CC: VA bond and hybrid to come soon, maybe VA can capture pitchfork bifurcation
%with enough VA parameters taken}]}
Finally we apply the methods to 3D
lattice solitons using the following site-centered ansatz
\begin{equation}
u_{m,n,l}^{\mathrm{(sc)}}=\left\{
\begin{array}{ll}
\beta  & \mathrm{~if~} m=n=l=0, \\[1ex]
Ae^{-\alpha (|m|+|n|+ |l|)} & \mathrm{~otherwise}
\end{array}%
\right.  \label{ansatzST3D}
\end{equation}%
where, for $C$ small enough, it also works well, see Fig.~\ref{3Dva1.ps}.

%By taking the 3D analogues of (\ref{Lag2}) and (\ref{ansatzST})--(\ref{ansatzHybrid}).

%%%%%%%%%%%%%%%%%%%%%%%%%%%%%%%%%%%%%%%%%%%%%%%%%%%%%%%%%
\section{Conclusion\label{Sec:conclusions}}

In this work, we have examined the existence, stability, and mobility of
discrete solitons in 2D and 3D NLS lattices with competing
(cubic-quintic, CQ) onsite nonlinearities. Some properties of the discrete
solitons, such as existence of the solutions of tall and short types,
each narrow and/or wide, resemble properties recently found in discrete
solitons in the 1D counterpart of this model \cite{Ricardo}, as well
as the 2D properties of models such as the one with the 
saturable nonlinearity \cite{Sweden}. We have found
pitchfork bifurcations connecting the site-centered and bond-centered
solitons via unstable asymmetric ones, in contrast with the 1D
model, where the connecting asymmetric solutions were stable. Another
fundamental soliton species that was studied in this work, \textit{viz}.,
hybrid solutions, exists only in the higher-dimensional lattice. We have
found, in some regions of the parameter space, that the site-centered and
bond-centered solitons were also connected via the hybrid states. At small
values of the inter-site coupling constant, $C$, various types of the 2D and
3D stationary discrete solitons are well described by the variational
approximation (VA).

We have also showed that enhanced mobility of 2D discrete solitons in the CQ
lattice can be realized by imparting to them kinetic energy exceeding the PN
barrier. Nevertheless, the moving solitons eventually come to a halt, due to
the radiation loss. In that connection, we were unable to find exact
transparency points at which translationally invariant solutions would be
able to exist. However, looking for carefully crafted radiationless
solutions for moving solitons in 2D and 3D lattice models remains a
challenging open problem. It would also be interesting to study the mobility of
the discrete solitons by means of a dynamical version of the the VA (in the
1D model with the cubic onsite nonlinearity, a dynamical VA was adapted to
the analysis of collisions between moving discrete solitons in Ref.~\cite{Papa}, and
to capture the stationary site-centered and bond-centered solutions
with a single ansatz in Ref.~\cite{kaup1}).

Getting back to stationary 2D and 3D discrete solitons in the cubic-quintic 
NLS lattice, remaining topics of interest
are to search for staggered solitons similarly 
e.g., to the work of Ref.~\cite{zc_06} for the cubic lattice, as well as lattice
solitons with intrinsic vorticity. Thus far, discrete lattice solitons 
and vortices were
studied in 2D \cite{vortex2D} and 3D DNLS equation with the cubic
nonlinearity \cite{vortex3D}.

%%%%%%%%%%%%%%%%%%%%%%%%%%%%%%%%%%%%%%%%%%%%%%%%%%%%%%%%%%%%%%%%%%%%%%%%%
\section*{Acknowledgments}
The authors would like to thank Guido Schneider, Dmitry Pelinovsky, and
Sergej Flach for insightful discussions.
The work of C.C.~was partially supported by the Deutsche Forschungsgemeinschaft DFG and the
Land Baden-W\"urttemberg through the Graduiertenkolleg GRK 1294/1:
Analysis, Simulation und Design nanotechnologischer Prozesse. %
R.C.G.\ acknowledges
support from NSF-DMS-0505663. P.G.K.\ acknowledges support from NSF-CAREER,
NSF-DMS-0505663 and NSF-DMS-0619492, as well as from the Alexander von
Humboldt Foundation. The work of B.A.M.~was in a part
supported by the Israel Science Foundation through the Center-of-Excellence
grant No. 8006/03.

%%%%%%%%%%%%%%%%%%%%%%%%%%%%%%%%%%%%%%%%%%%%%%%%%%%%%%%%%%%%%%%%%%%%%%%%%%

\end{document}